\documentclass[twocolumn,floatfix,prb,aps,showpacs,superscriptaddress]{revtex4-2}
\usepackage{graphicx,amsmath,amssymb,color}
\usepackage[utf8]{inputenc}
\usepackage{nicefrac}
\usepackage{multirow, makecell, ragged2e}
\usepackage[titletoc,title]{appendix}

\usepackage{color,soul}
\usepackage[dvipsnames]{xcolor}
\definecolor{DarkGreen}{rgb}{0.0, 0.5, 0.0}
\usepackage[colorlinks,citecolor=DarkGreen,linkcolor=blue,linktocpage]{hyperref}

\newcommand{\be}{\begin{equation}}
\newcommand{\ee}{\end{equation}}

\newcommand{\ba}{\begin{eqnarray}}
\newcommand{\ea}{\end{eqnarray}}

\begin{document}

\title{Non-equilibrium physics of the density-difference-dependent Hamiltonian: quantum scarring in the presence of dynamical gauge fields and chiral symmetry}
\author{W. N. Faugno}
\affiliation{Laboratoire Kastler Brossel, Coll\`ege de France, CNRS, ENS-Universit\'e PSL, Sorbonne, Universit\'e, 11 Place Marcelin Berthelot, 75005 Paris, France}
\author{Hosho Katsura}
\affiliation{Department of Physics, University of Tokyo, Hongo, Bunkyo-ku, Tokyo 113-0033, Japan}
\affiliation{Institute for Physics of Intelligence, University of Tokyo, Hongo, Bunkyo-ku, Tokyo 113-0033, Japan}
\affiliation{Trans-Scale Quantum Science Institute, University of Tokyo, Bunkyo-ku, Tokyo 113-0033, Japan}
\author{Tomoki Ozawa}
\affiliation{Advanced Institute for Materials Research (WPI-AIMR), Tohoku University, Sendai 980-8577, Japan}
\date{\today}

\begin{abstract}
Quantum many-body scars represent a form of weak ergodicity breaking that highlights the unusual physics of thermalization in quantum systems. Understanding scar formation promises insight into the connection between classical statistical mechanics and the quantum world. The existence of quantum many-body scars calls into question how the macroscopic world can arise from the Schr{\"o}dinger equation. In this work, we demonstrate the existence of quantum many-body scars in the density-difference-dependent Hamiltonian. This Hamiltonian has a particular manifestation of chiral symmetry due to its interaction being neither attractive nor repulsive a priori, but depending on the configuration. As a result of this symmetry and peculiar interaction, we find that this system hosts two different classes of quantum scars; a charge density wave ordered scar and an edge-mode scar. We establish the existence of these scars by examining the entanglement entropy of the system as well as demonstrating robust thermalization breaking time dynamics. For each, we propose simple mechanisms that give rise to these scars which may be applicable to other systems. 
\end{abstract}

\maketitle

\section{Introduction}
The unitary dynamics of quantum mechanics and the ergodicity assumed in classical statistical mechanics appear at first to be incompatible. Unitary dynamics imposes that the infinite time-averaged value of observables will be the weighted sum of the diagonal contributions in the observable. As such, the system appears to be non-ergodic and does not follow a trajectory that occupies all configurations within a given energy shell. If quantum mechanics is a more fundamental theory, how can classical statistical mechanics rely on an assumption that contradicts the basic phenomenon of unitary evolution? A potential remedy to this seeming contradiction is found in the eigenstate thermalization hypothesis (ETH)~\cite{Jensen85,Deutsch91,Srednicki94,Rigol08}. The ETH states that the thermodynamic expectations of an observable in a quantum system will agree with the microcanonical ensemble so long as the expectations of the observable in an eigenstate vary smoothly with energy and off-diagonal contributions vanish exponentially with system size. The ETH has been generally successful in connecting non-integrable quantum systems with the expectations of classical statistical mechanics~\cite{DAlessio16}.

Despite the success of the ETH, systems have been found which host violations where seemingly chaotic quantum systems do not always thermalize. The most robust violation of the ETH is many-body localization (MBL) where all eigenstates become non-thermalizing despite the presence of an interaction~\cite{DeLuca13,Levi16}. The lack of thermalization here can be understood as the system displaying an emergent integrability where an extensive set of local integrals of motion (LIOMs) emerge~\cite{Serbyn13b,Huse14}. Recent efforts have indeed found that these LIOMs can be constructed explicitly in the MBL phase in several contexts~\cite{Chandran15,Rademaker16,Thomson18,Pekker17}. Currently, it is debated whether the MBL phase is a true phase transition surviving in the thermodynamic limit due to rare regions of low disorder destabilizing the localized phases~\cite{DeRoeck17,Thiery18}. Still, evidence for MBL has been established in finite systems~\cite{Kohlert19,Morong21,Guo21,Guo23b} and it survives as a dynamical phase if not a thermodynamic phase~\cite{Pal10}. The behavior displayed by MBL is termed strong ETH violation as all eigenstates of the system are non-thermal despite the initial Hamiltonian appearing chaotic.

Weak ETH violations, i.e. ETH violations by only a small subset of eigenstates, have also been observed in the form of quantum many-body scars (QMBS)~\cite{Serbyn21,Moudgalya22,Chandran23}. These few eigenstates which make up the QMBS promote the ``memory" of specific initial conditions in quantum systems, in contrast to MBL where any initial condition is preserved. These special initial states do not decohere as expected for a generic quantum state, as shown by revivals in the time-dependent fidelity. A paradigmatic model featuring QMBS is the PXP model, which describes a Rydberg atom chain in the blockade limit~\cite{Serbyn21}. The PXP model has multiple families of scars, but the most prominent is the so-called $\mathbb{Z}_2$ family of scars, named for their high overlap with a state with antiferromagnetic/charge density wave (CDW) order. These states exhibit low entanglement entropy and anomalously large average spin polarization. From an initial antiferromagnetically ordered state, the system shows periodic sharp revivals in the time-dependent fidelity, with the frequency determined by the energy difference between scar eigenstates~\cite{Turner18,Turner18a,Turner18b,Turner21,Hudomal22,Ljubotina23}. As yet the mechanism for QMBS formation in the PXP model is not fully understood, but various investigations have made promising progress~\cite{Moudgalya20,Huang21b,Omiya23,Buijsman24}.

An important aspect of the PXP Hamiltonian in QMBS formation comes from the kinetic constraint imposed by the projector. This has inspired a number of models with kinetic constraints, including bosonic and fermionic, that have been proposed to realize different forms of QMBS~\cite{Pancotti20,Hudomal20,Tamura22,gotta2022exact,Sanada23,Su23,Keneko24}. QMBS have also been proposed in other scenarios, arising from geometric frustration~\cite{McClarty20}, conserved quantities~\cite{Sala20}, truncated Hilbert space, and dynamical constraints~\cite{Lan18}. Some QMBS have also been associated with unstable classical orbitals~\cite{Hummel23,Evrard24}. QMBS are often found in systems with Hilbert space fragmentation where the Hilbert space separates into exponentially many subspaces that are dynamically disconnected.~\cite{Yang20,Moudgalya22,Frey24} They may also be characterized by the complexity of the Krylov subspace generated by the Hamiltonian, which better captures systems where fragmentation is not exact~\cite{Bhattacharjee22,Nandy24}. Further, it has been demonstrated that the nonergodic behavior of QMBS may survive even when the scar states are not exact eigenstates, but are stabilized in some limit of the model, a phenomenon known as asymptotic QMBS~\cite{Gotta23,Kunimi24}.

In this work, we present two new mechanisms for QMBS formation in the density-difference-dependent Hamiltonian, which to our knowledge do not rely on the previously known mechanisms mentioned above. This model was previously proposed in the context of Floquet engineering and has been explored in the classical limit~\cite{Faugno24} and non-Hermitian few body limit~\cite{Faugno23}. In both previous works, we found that chiral symmetry played a key role in understanding the system. 
Now we consider this model for a system of bosons at density $\nu = \frac{N}{L} = \frac{1}{2}$, where $N$ is the total number of bosons and $L$ is the total number of lattice sites. We find that there are two new mechanisms for stabilizing QMBS, leading to two kinds of scars.
The first QMBS is associated with a charge density wave. This state is stabilized by the destructive interference of the bare hopping and the correlated hopping (equivalently, gauge field hopping). As we will see this is not a true QMBS, but is a form of weak scarring with the non-ergodic fidelity oscillations vanishing in the thermodynamic limit. The second type of QMBS we discuss is associated with initial states of many-body edge modes. As we will see, these QMBS are stabilized by the interplay of energy detuning between edge states, which arise as truncation-induced edge states in an effective Fock space lattice picture, and the rest of the spectrum, and the chiral symmetry of our model.

\section{The Density-Difference-Dependent Hamiltonian}
We consider bosons on a one-dimensional chain of length $L$ described by 
the density-difference-dependent Hamiltonian
\begin{align}
    H = & \sum_j a^\dagger_{j+1}[-J+\gamma(n_{j+1}-n_j)]a_j \nonumber \\
    & + a^\dagger_{j}[-J+\gamma^*(n_{j+1}-n_j)]a_{j+1},
    \label{eq:Ham}
\end{align}
where $a_j^\dagger$ and $a_j$ are bosonic creation and annihilation operators, respectively, $J$ is the hopping parameter, and $\gamma$ is the coupling to the density-difference-dependent hopping which we allow to be complex in general. We will set $J=1$ for the rest of this article unless otherwise stated. In this work, we will consider both periodic boundary conditions (PBC) and open boundary conditions (OBC). This model was previously proposed to be realizable as the effective Floquet Hamiltonian of a bosonic system under a three-step periodic drive~\cite{Faugno23}. From this Floquet method, $\gamma$ is most naturally purely imaginary, but can be made real or complex in principle. The density-dependent gauge can be interpreted in multiple ways. This includes a dynamical gauge field, an interaction, and a correlated hopping. This correlated hopping interpretation is precisely the motivation for looking for quantum scars, as many other proposed scars result from similar density-dependent hopping~\cite{Hudomal20}. Note that for imaginary $\gamma$, the Hamiltonian is symmetric under the parity transformation, permitting a block diagonal form in two blocks. The parity transformation is inversion about the center of the chain, i.e., for a chain with site labels $j=1, ..., L$, the transformation is $a_j^{(\dagger)}\rightarrow a_{L+1-j}^{(\dagger)}$. In general, the exact form of the transformation depends on the choice of origin. 

This Hamiltonian has a 
chiral symmetry, $a_j^{(\dagger)}\rightarrow (-1)^ja_j^{(\dagger)}$, which transforms $H$ to $-H$, imposing that for each eigenstate with energy $E$ there exists one at $-E$. The chiral symmetry can equivalently be represented by labeling the Fock states with a dipole moment which we define as $d=(-1)^{\sum_j j n_j}$. In actuality, this is the parity of the more typically defined dipole moment, $\sum_j j n_j$, but we adopt this definition as it is well defined in both PBC and OBC~\cite{Lake22,Lake23,Oh24}. The Hamiltonian can be brought to a block off-diagonal form in two blocks where each corresponds to either parity of the dipole. The dimensions of the off-diagonal blocks are in general not the same. The difference between the dimensions of the dipole sectors guarantees the existence of at least that many zero energy eigenstates, protected by the chiral symmetry.
We obtain a formula for the minimal number of zero energy states by partitioning the lattice sites into odd and even sites. Then a configuration where an odd number of particles lie on the odd sites has dipole $-1$ while an even number of particles on the odd sites has dipole $1$. Taking the difference between the number of states in each parity sector, for a system of $N$ particles on $L$ sites, we obtain the formula
\begin{equation}
    \mathcal{N}_{E=0} = (-1)^{NL}\sum_{n=0}^N (-1)^n \binom{
        \lceil \frac{L}{2} \rceil - 1 + n
        }{n}
    \binom{
        \lfloor \frac{L}{2} \rfloor - 1 + N - n
        }{N-n}
\end{equation}
for the lower bound on the number of zero energy states protected by the
chiral symmetry. This is a generic result for any Hamiltonian with this dipole structure and has been reported in other recent works~\cite{Theel25}.

Before searching for quantum scars, we first demonstrate that this model is chaotic by examining the average level spacing ratio. Ordering the energies from lowest to highest, we define the gaps as $S_n = E_n-E_{n-1}$. Then the level spacing ratio is defined as $r_n = \min(S_n,S_{n-1})/\max(S_n,S_{n-1})$. If a system is chaotic, the average level spacing ratio is expected to match that of Gaussian random ensembles, obtained from random matrix theory (RMT), with the specific ensemble depending on the time-reversal symmetry of the system. If the Hamiltonian has time-reversal symmetry, one should compare with the Gaussian orthogonal ensemble (GOE), while if time-reversal symmetry is broken, one should compare with the Gaussian unitary ensemble (GUE)~\cite{Atas13}. Additionally, in the presence of symmetries, one should compare with the appropriate Gaussian ensemble with block diagonal form corresponding to the degree of the symmetry, $m$~\cite{Giraud22}. For purely real $\gamma$, our Hamiltonian is real and thus time-reversal symmetric and we compare with the GOE. For purely imaginary $\gamma$, our system breaks time reversal symmetry with a two-fold symmetry coming from the parity symmetry mentioned above and so we compare with the GUE consisting of two block diagonal submatrices.

\begin{figure}
    \centering
    \includegraphics[width=\linewidth]{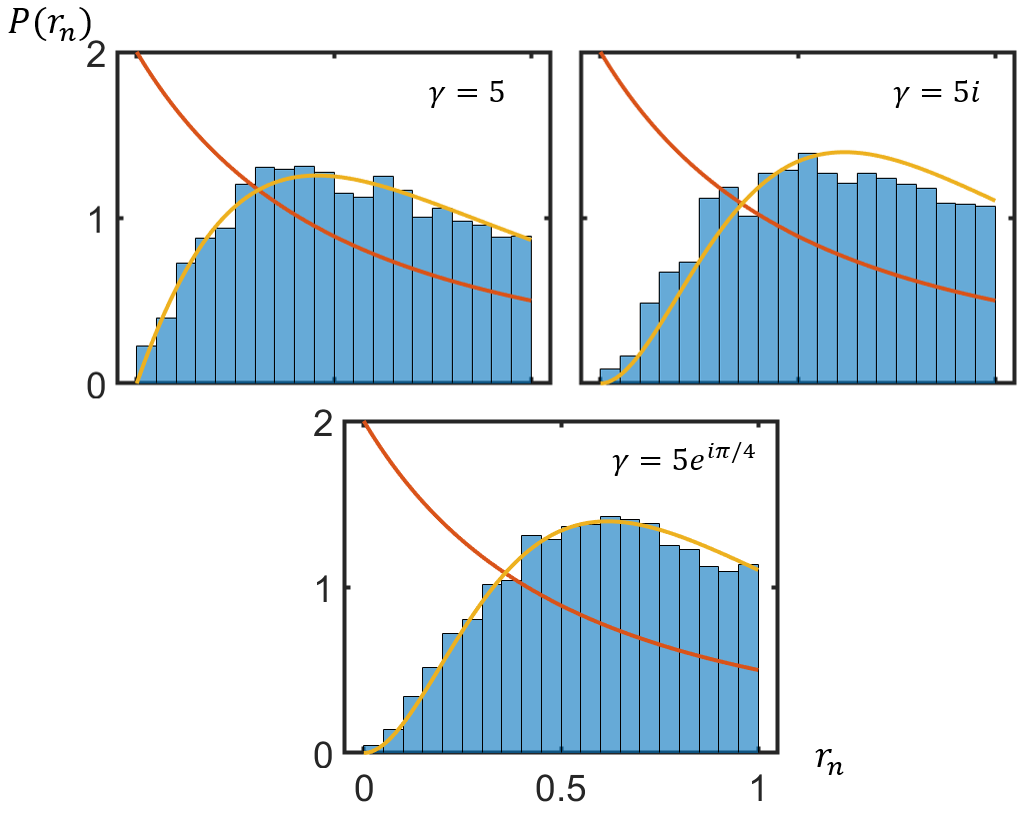}
    \caption{Probability density obtained from histogram of level statistics for real, imaginary, and complex $\gamma$ under OBC. These results are for $N=6$, $L=12$. We plot the expected probability densities for an integrable system (Poissonian, orange) and chaotic (Wigner-Dyson, yellow), which demonstrates that the system is chaotic. The parameters for the Wigner-Dyson distributions depend on the random matrix ensemble under consideration (see~\cite{Giraud22}). Note that for $\gamma=5i$, purely imaginary $\gamma$, the distribution is only for one parity sector.}
    \label{fig:WignerDyson}
\end{figure}

In OBC, the level spacing ratio average for our model at half-filling and length $L=12$ (resulting in a Hilbert space dimension of 12376) with real, imaginary, and complex ($\gamma = |\gamma|e^{i\pi/4}$) $\gamma$ are as follows for $|\gamma|=5$: $0.5292$ (GOE, $0.53590$), $0.4292$ (GUE $m=2$, $0.422085$), and $0.6079$ (GUE, $0.60266$). We have included the corresponding Gaussian ensemble and the expected results in parentheses. For imaginary $\gamma =5i$, We also look at the level statistics within one of the parity blocks for purely imaginary $\gamma$ and obtain $0.5723$ close to the GUE prediction. Further evidence is provided by looking at the probability density, $P(r_n)$, provided in Fig.~\ref{fig:WignerDyson} where we see that the level statistics clearly follow a Wigner-Dyson distribution (yellow curve) as opposed to a Poissonian (orange curve) expected of integrable systems. Note that for $\gamma =5i$, we only present one of the parity sectors. Now that we have established that the system is chaotic, we present the two mechanisms for scar formation. First we consider $\gamma$ to be purely imaginary, resulting in scarring that stabilizes a CDW-like mode. Then we consider $\gamma$ to be purely real, which results in scar eigenstates that support many-body edge modes.

\section{Charge Density Wave Scar}
\begin{figure*}
    \centering
    \includegraphics[width=\linewidth]{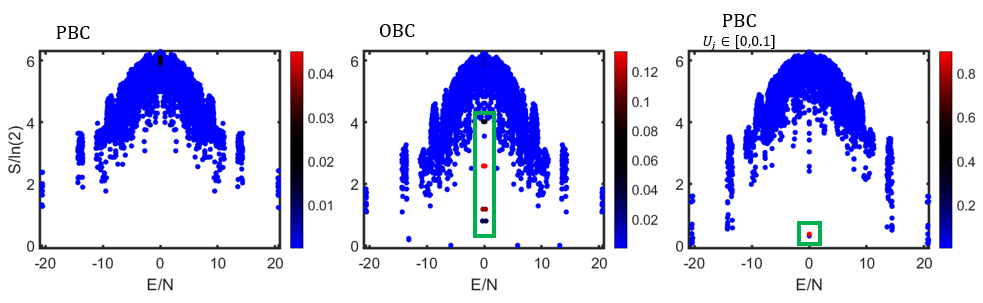}
    \caption{Entanglement entropy spectrum with $\gamma=5i$ for a system of $N=6$, $L=12$ for PBC (left), OBC (center), and PBC with random onsite interaction (right). The states within the green box on the middle and right panels are the eigenstates that stabilize the scar dynamics. The coloring is given by the overlap of the eigenstate with $|\Psi_0\rangle=|{\rm CDW}\rangle$.}
    \label{fig:imGamEE}
\end{figure*}

Consider $\gamma \in \mathbb{R}i$, that is $\gamma$ to be purely imaginary. To establish the presence of QMBS, we partition the lattice in real space into two halves, labeling them subsystems $A$ and $B$, and calculate the bipartite entanglement entropy as $S=-\sum \lambda_A^2 \log \lambda_A^2$ where $\lambda_A^2$ are the eigenvalues of the reduced density matrix of subsystem $A$ after tracing over subsystem $B$, looking for states with anomalously low entanglement entropy, suggestive of ETH violation. We present the entanglement spectrum for $N=6$, $L=12$ and $\gamma = 5i$ under PBC (left) and OBC (center) in Fig.~\ref{fig:imGamEE}. In PBC, the spectrum does not show any sign of QMBS, but this is due to the chiral and parity symmetries stabilizing a large number of zero energy modes that the QMBS state mixes with. On the other hand, in OBC we observe several low entanglement entropy states. Those at zero entanglement are edge modes which we will explore for the case of real $\gamma$ later, and thus we will ignore them here. Here we focus on those low-entanglement eigenstates near zero energy boxed in green. These have a relatively high overlap with both CDW order states $|{\rm CDW}\rangle = \prod_{j=1}^{L/2} a^\dagger_{2j-1}|0\rangle$ and $|{\rm CDW'}\rangle = \prod_{j=1}^{L/2} a^\dagger_{2j}|0\rangle$, where $|0\rangle$ is the vacuum state with no bosons. To show that these scar states are still present in PBC, we add a random onsite interaction term of the form $\sum_j U_jn_j(n_j-1)$ where $U_j\in[0,0.1]$ is uniformly distributed. This lifts the zero energy degeneracies by breaking the chiral symmetry. We plot the entanglement entropy spectrum in the presence of disorder in the right panel of Fig.~\ref{fig:imGamEE}. Here we observe two low entanglement states corresponding to the CDW and CDW$'$ orders, boxed in green. (Note that these results are for a single disorder realization and not averaged.)

\begin{figure}
    \centering
    \includegraphics[width=\linewidth]{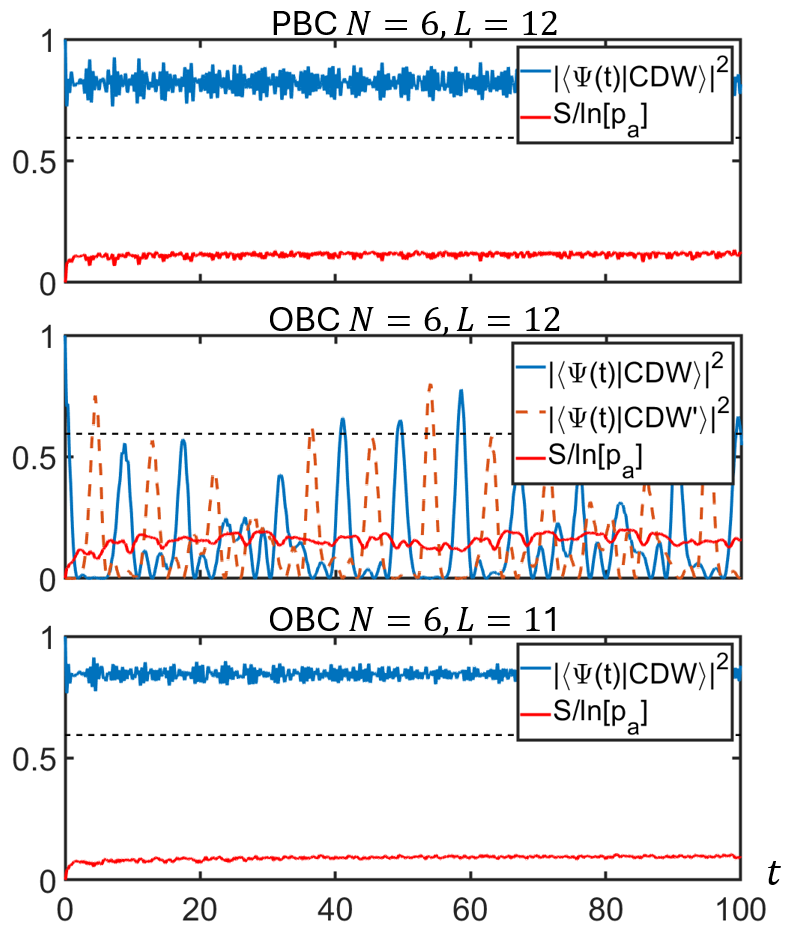}
    \caption{Time dependent fidelity and entanglement entropy growth for an initial state $|\Psi_0\rangle = |{\rm CDW}\rangle$. Here we set $J=1,\gamma=5i$ for a system of $N=6,L=12$. The upper panel shows the results for PBC where the CDW order is stable and the entanglement entropy saturates quickly. The middle panel shows the results for OBC where the fidelity and entanglement entropy oscillate. We also show the fidelity with the CDW' state to demonstrate how the system oscillates between the two orders. In the lower panel, we change the length to $L=11$ and show the fidelity under OBC where the fidelity and entanglement entropy show the same behavior as in PBC. The dashed line corresponds to the page value where we expect the entanglement entropy to saturate for a thermal state.}
    \label{fig:imGamTime}
\end{figure}

These high overlaps motivate us to look for nonergodic dynamics from the initial configuration $|{\rm CDW}\rangle$. We present the time evolution of the fidelity and entanglement entropy in Fig.~\ref{fig:imGamTime} for PBC (upper) and OBC (middle). We find that the fidelity with the CDW state is remarkably stable in PBC and does not drop to zero, but remains large in proportion to the value of $|\gamma|$. Under OBC, the fidelity shows dynamics with oscillations typical of QMBS. In both cases, the entanglement entropy growth is strongly suppressed. For an ergodic state, we would expect the entanglement entropy to quickly grow and saturate at the Page value as given by the black dashed line in the figure~\cite{Page93,Vidmar17,Bianchi19,Bianchi22}. We also present the time-dependent fidelity with $|{\rm CDW}'\rangle$ for OBC, which shows that the system slides between the two CDW orders. The frequency of oscillation is related to the energy separation of the scar eigenstates. Since these states are not equally separated in energy, there are multiple frequencies in the dynamics corresponding to the different separations. We have found that these frequencies are proportional to the hopping $J$ and independent of the value of $\gamma$ so long as it is large enough that the scars are stabilized. The frequency appears to be inversely proportional to the length of the chain $L$ as well. We provide the numerical evidence for both the $J$ dependence and system size dependence in Appendix \ref{app:frequency}.

In the lower panel of Fig.~\ref{fig:imGamTime}, we present the fidelity and entanglement entropy for an OBC chain with $N=6$, $L=11$ where the CDW terminates with an occupied site on both edges ($|\text{CDW}\rangle=\prod_{j=1}^{6}a_{2j-1}|0\rangle$). We observe that the behavior is very similar to the PBC case, suggesting that the transfer between the CDW orders observed in OBC is a result of the unoccupied edge. We will see the reason for this dependence on the parity of the chain length when we describe the mechanism responsible for the scarring.

\begin{figure}
    \centering
    \includegraphics[width=\linewidth]{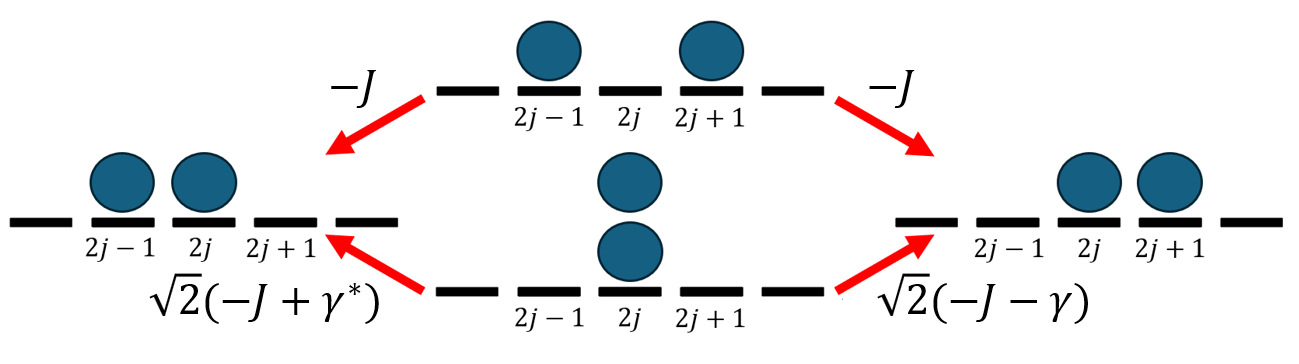}
    \caption{Diagram of the hopping processes that destructively interfere to stabilize the CDW scar states.}
    \label{fig:hopDia}
\end{figure}

We have found that the scar eigenstates are the result of destructive interference between the bare hopping and correlated hoppings. We diagram the scenario in Fig.~\ref{fig:hopDia}. To stabilize the CDW, the system must cancel the bare hopping given by $-J$ present in the CDW state. On the $2j$th site, this is achieved by mixing with a state of the form $(a^\dagger_{2j})^2a_{2j-1}a_{2j+1}|{\rm CDW}\rangle$ where the hopping to unbind particles costs $\sqrt{2}(-J-\gamma)$ for rightward hoppings and $\sqrt{2}(-J+\gamma^*)$ for leftward hoppings. Then the CDW state can be stabilized by taking an appropriate superposition of these two configurations. As such, the scar states can be approximated by 
\begin{align}
|\Psi_{\rm scar}\rangle = c_0|{\rm CDW}\rangle &+ \sum^{L/2}_{j=1} c_j(a^\dagger_{2j})^2a_{2j+1}a_{2j-1}|{\rm CDW}\rangle,\\
\frac{c_j}{c_0} &= \frac{-J}{\sqrt{2}(J+\gamma)}=\frac{-J}{\sqrt{2}(J-\gamma^*)}=\alpha\label{eq:AnsatsCoeff}
\end{align}
From Eq.~(\ref{eq:AnsatsCoeff}), we see that for both hopping directions to be suppressed, the coupling $\gamma$ must be purely imaginary. This ansatz is accurate when $\gamma/J$ is large as higher order terms are proportional to $|J/\gamma|^n$, involving $n$ bindings or unbindings of particles. At the next order, terms appear with weight at $|J/(J+\gamma)|^2$ from two processes, which we explain in more detail in Appendix \ref{app:next_order}.

\begin{figure}
    \centering
    \includegraphics[width=\linewidth]{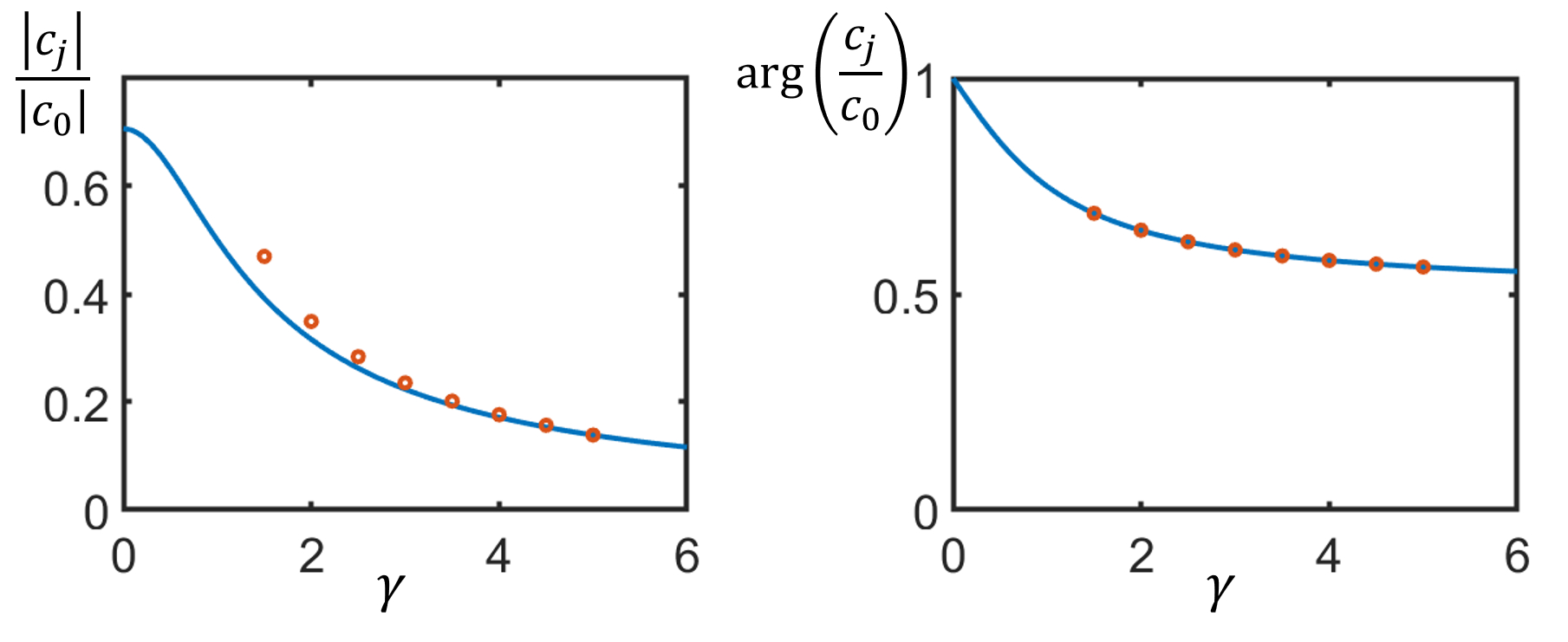}
    \caption{Ratio of the coefficients (left) and relative phase (right) of the ansatz for the scar state. The curve is the analytical result obtained from Eq.~(\ref{eq:AnsatsCoeff}) while the points are the numerical results for $L=10$, $N=5$ in PBC.}
    \label{fig:CDWrat}
\end{figure}

We demonstrate that our eigenstate approximates the scar well for a system with $N=5$, $L=10$ at large $\gamma$ even down to $\gamma = 1.5i$ in Fig.~\ref{fig:CDWrat} where we plot the numerically obtained ratio of the coefficients $|c_j|/|c_0|$ (left) and relative phase (right) alongside the analytical curves. We obtain the scar by exact diagonalization of the system under PBC. (Note we use $L=10$ because in systems with $N$ even there exists a large number of zero energy eigenstates that mix with this simple state hiding its structure. This degeneracy can be broken for even $N$ by going to OBC where we observe an approximate form of the scar state, but here we have opted to consider the smaller system with an odd number of particles.) Thus we have found the primary mechanism behind the CDW-like scar formation. This mechanism fails when $\gamma$ has a real part as the unbinding hopping process of the two particle state has different energy costs for right and left directions.

\begin{figure}
    \centering
    \includegraphics[width=\linewidth]{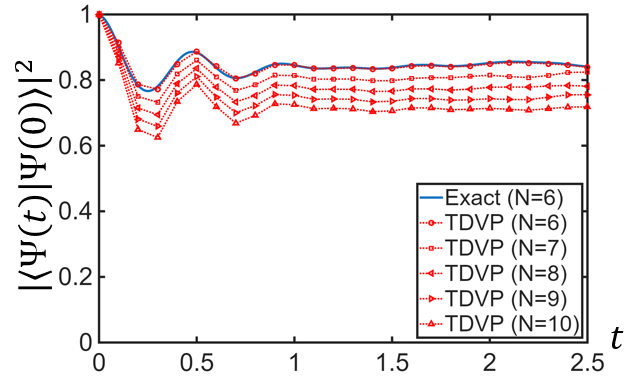}
    \caption{System-size dependence of the time-dependent fidelity for $\gamma=5i$. Here we employ TDVP to show that as the system size increases, the fidelity saturates at a smaller value. We consider systems of $N$ particles on a chain of length $L=2N-1$. The initial configuration is a CDW that terminates in an occupied site on both ends of the chain.}
    \label{fig:TDVPcdw}
\end{figure}

We can further demonstrate how the fidelity decays for these scars in the thermodynamic limit. If we combine the relation in Eq.~\ref{eq:AnsatsCoeff} with the normalization condition we find that
\begin{equation}
    |c_0|^2+\frac{L}{2}|c_j|^2 = 1
\end{equation}
which gives the scaling for $c_0$ as
\begin{equation}
    |c_0|^2 = \frac{2|\alpha|^2}{L+2|\alpha|^2}
\end{equation}
which shows that the weight of the scarred configuration decreases with system size $L$ and vanishes in the thermodynamic limit. We employ a time-dependent variational principle calculation (TDVP) using the ITensor package~~\cite{10.21468/SciPostPhysCodeb.4,10.21468/SciPostPhysCodeb.4-r0.3} to demonstrate that the time-dependent fidelity decreases as the system size increases as shown in Fig.~\ref{fig:TDVPcdw}. TDVP is a method for efficiently approximating the dynamics of a quantum system by restricting the evolution to a sub-manifold of Hilbert space. Thus it works well when the dynamics remains close to this manifold making, TDVP a natural method for studying scar dynamics. Here we show the time-dependent fidelity for systems of $N$ particles on a chain of length $L=2N-1$ where we expect the fidelity from an initial CDW configuration to quickly saturate. As we observe, the saturation value decreases with increasing system size. This suggests that this is not a true QMBS, but a form of weak scarring. Note that we have not considered the case where there is an open boundary as the period of oscillations is too long for us to observe the first revival in our TDVP calculation.

This mechanism explains the transfer between the two CDW orders in OBC when the length $L$ is even. At an unoccupied edge, e.g. $|1010...10\rangle$, the particle is free to hop toward the edge as it is not possible to form the necessary doublon to enforce the destructive interference condition. After this particle moves towards the edge, the next particle in the chain is free to move as well, resulting in a cascade where particles transfer down the chain one at a time only controlled by the bare hopping. This explains the $J$ and $L$ dependence of the frequency as well as the $\gamma$ independence. Interestingly, this suggests that there are different thermodynamic behaviors depending on the parity of the chain one considers. This leads us to conclude that the scar may be unstable, especially to deviations in the density. Further evidence for the existence of the CDW scar is provided in Appendix \ref{app:kyrlov} through a Krylov subspace analysis.

\section{Many-Body Edge Mode Scar}
\begin{figure}
    \centering
    \includegraphics[width=\linewidth]{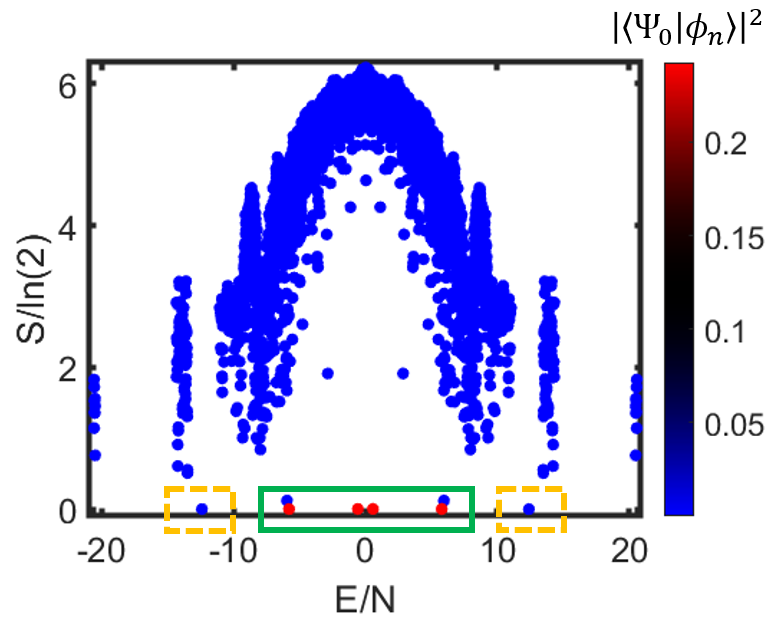}
    \caption{Entanglement entropy spectrum for $N=6$, $L =12$, and $\gamma=5$. Points are colored according to the overlap between the corresponding eigenstate $|\phi_n\rangle$ and the state $|\Psi_0\rangle= (a^\dagger_1)^3 (a^\dagger_L)^3|0\rangle$. The states boxed in green are those that stabilize the edge mode scar, while those within the yellow dashed box are not scars, but many-body edge modes.}
    \label{fig:rGam_EE_6_12}
\end{figure}

We now consider the case of real $\gamma$ at half-filling in the full many-body system. We calculate the bipartite entanglement entropy for a half-filled system to find ETH violating states. We present the entanglement entropy for a system with $N=6$ and $L=12$ and coupling $\gamma= 5$ in Fig.~\ref{fig:rGam_EE_6_12}. We will refer to the central continuum of states as the ``high-temperature'' region and the separated continuums as the ``bound spectra.'' There are several states which clearly violate the ETH, having zero entanglement entropy, which we have found to be edge states. We separate these into two classes where those boxed in green correspond to an edge mode that occupies both edges, while those in the yellow dashed box only occupy one edge. (Both states within the yellow dashed box localize on the same edge, determined by the sign of $\gamma$. States localized on the opposing edge exist, but lie within the bound spectra for this choice of parameters.) The exact edge mode configuration depends on the particle number. For example, $(a^\dagger_1)^5a^\dagger_2|0\rangle$ is more stable than $(a^\dagger_1)^6|0\rangle$. We consider those in the yellow dashed box to be out of the high-temperature part of the spectrum, and thus will not consider them as QMBS. They are the result of energy detuning between edge states and the rest of the spectrum, as has been analogously observed in the Bose-Hubbard model through perturbation theory~\cite{Pinto09}. This mechanism was shown to stabilize many-body edge modes for systems of more than three particles in the Bose-Hubbard model when the interaction dominates. For our system, the edge states are found to be detuned from the rest of the bound spectra in the limit that $J\rightarrow 0$ and can be understood as truncation-induced edge states as we will discuss below. Even though we do not consider these states as QMBS at $N=6$, their energy detuning can combine with the emergent chiral symmetry to stabilize QMBS for a system of $N=12$ as we discuss below.

\begin{figure}
    \centering
    \includegraphics[width = \linewidth]{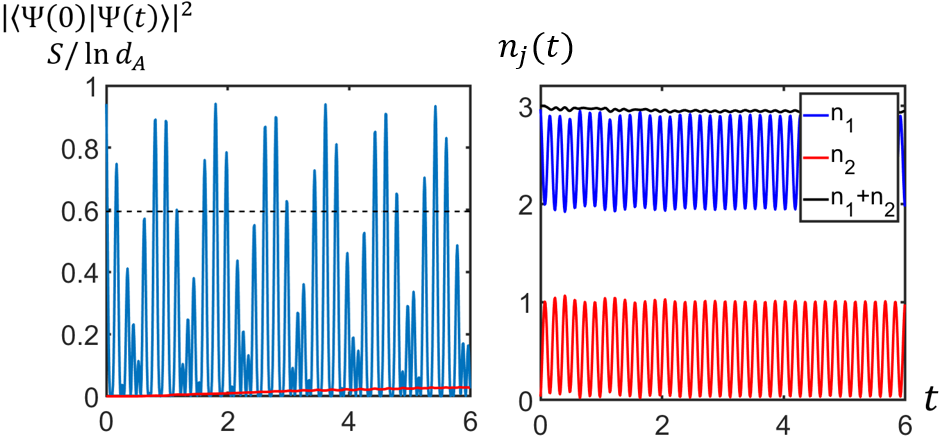}
    \caption{Nonergodic dynamics of the initial state $|\Psi_0\rangle = (a^\dagger_1)^3 (a^\dagger_L)^3|0\rangle$. The system consists of $N=6$ particles on a chain of length $L=12$ with $J=1$, $\gamma=5$. (Left) We plot the time-dependent fidelity taken between the initial state and the state evolved after a time $t$ in blue and the entanglement entropy growth in red. The black dashed line corresponds to expected entanglement entropy saturation of an ergodic system. (Right) The densities on the first and second sites as well as their sum demonstrating the simple dynamics of the system when initialized in this configuration.}
    \label{fig:rGamTimeEvo}
\end{figure}

The states boxed in green lie within the thermalizing portion of the spectrum and are thus QMBS candidates. Note that these states have a high overlap with the simple product state $|\Psi_0\rangle = (a^\dagger_1)^3 (a^\dagger_L)^3|0\rangle$ as indicated by the coloring of points in Fig.~\ref{fig:rGam_EE_6_12}. This leads us to test the dynamics from the initial state $|\Psi_0\rangle$ for nonergodic behavior. We plot the fidelity of the initial state $|\Psi_0\rangle$ as a function of time in the left panel of Fig.~\ref{fig:rGamTimeEvo}, which shows the telltale revivals for QMBS. Additionally, we plot the entanglement entropy normalized by $\ln d_A$ where $d_A$ is the dimension of the density matrix after tracing over subsystem $B$, as a function of time and see that its growth is strongly suppressed well below the Page value (black dashed line), again demonstrating the nonergodic nature of the dynamics. In the right panel, we plot the density of the first (blue) and second (red) sites along with their sum (black). The density shows that the particles remain stuck to their edge as a particle hops back and forth.

\begin{figure}
    \centering
    \includegraphics[width=\linewidth]{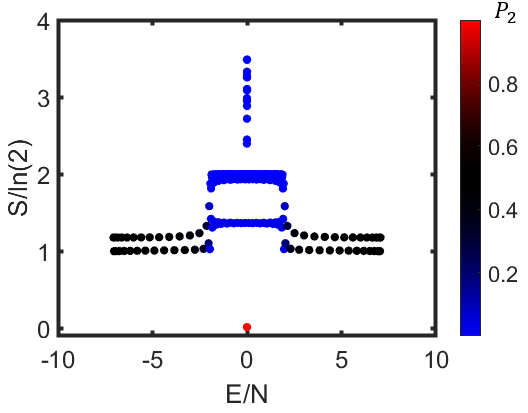}
    \caption{Entanglement entropy spectrum of the system with $N=2$ and $L=30$ and $\gamma=5$. The points are colored according to the expectation of $P_2 = \sum_j a_j^\dagger a_j^\dagger a_ja_j$.}
    \label{fig:2partEEgam}
\end{figure}

\begin{figure}
    \centering
    \includegraphics[width=\linewidth]{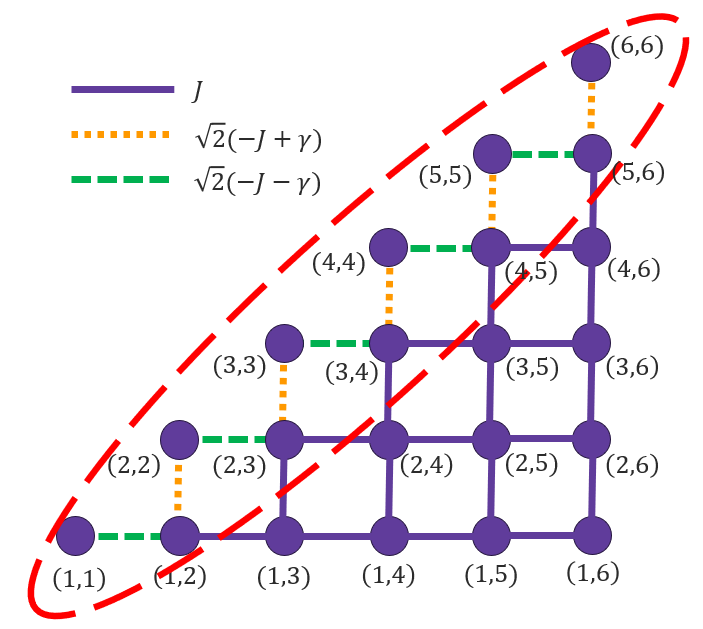}
    \caption{Lattice in Fock space for $N=2$, $L=6$. Labels correspond to the positions of the two particles, i.e. $(j,k) \equiv a^\dagger_ja^\dagger_k|0\rangle$. Bonds are colored according to the hopping amplitude obtained from the Hamiltonian in Eq.~(\ref{eq:Ham}). The dashed red oval encircles the states that correspond to an effective SSH model of bound pairs of particles.}
    \label{fig:Fock}
\end{figure}

The origin of these edge states lies in an effective Fock-space lattice description of clustered states of $N/2$ bosons combined with the chiral symmetry. To understand this mechanism, it is instructive to first consider the case of a two-particle system under real coupling, $\gamma$. In Fig.~\ref{fig:2partEEgam}, we present the entanglement entropy spectrum for a system of two particles, $N=2$, on a chain of length $L=30$ sites with $\gamma=5$. We observe that at zero energy there exists a state with almost zero entanglement. The coloring on the plot corresponds to the expectation of $P_2 = \sum_j a_j^\dagger a_j^\dagger a_ja_j$, which measures the degree of particle binding. The state at zero energy demonstrates near perfect binding. This state is, in fact, a two-particle edge mode that exists in analogy with the topological edge mode observed in a Su-Schrieffer-Heeger (SSH) model. This can be made explicit by constructing an effective lattice with chiral symmetry in Fock space where the sites in the A sublattice are given by states with particles on the same site, $a^\dagger_{A,j} = (a_j^\dagger)^2$, and the B sublattice sites correspond to configurations where particles lie on adjacent sites, $a^\dagger_{B,j} = a_j^\dagger a^\dagger_{j+1}$~\cite{Faugno23}. We diagram the Fock space lattice for $N=2$, $L=6$ in Fig.~\ref{fig:Fock}, with the subsystem that corresponds to the effective SSH model within the red dashed oval. Here we observe an emergent chiral symmetry, which gives rise to the SSH-like physics. An analogous phenomenon was reported for photons in coupled resonators \cite{Gorlach18}.

The entanglement entropy for two particles, presented in Fig.~\ref{fig:2partEEgam} has a high degree of structure beyond the zero energy edge state. First observe the two bands of states with a large expectation of $P_2$ (about $50\%$) and an entanglement entropy around $\ln{2}$. These are actually the extended states of the effective SSH model that the bound states obey. Since the bound states behave as a single particle here, the entanglement entropy is equivalent to that of a single particle. The slightly offset states are the results of the bound state being cut when we partition the system and the difference in entanglement entropy between these sets of states should vanish in the thermodynamic limit. The bands where the binding is minimal, (i.e. $P_2$ is small) show the same behavior as a system of two non-interacting bosons with some slight corrections due to the presence of the gauge field. This is the behavior for large $\gamma$ where there is a separation between two classes of states: those where particles avoid each other and those where particles lie on the same or adjacent sites. In the limit of $J=0$ this mapping is exact for two particles, but there is no gap in the doublon sector.

\begin{figure*}
    \centering
    \includegraphics[width=\linewidth]{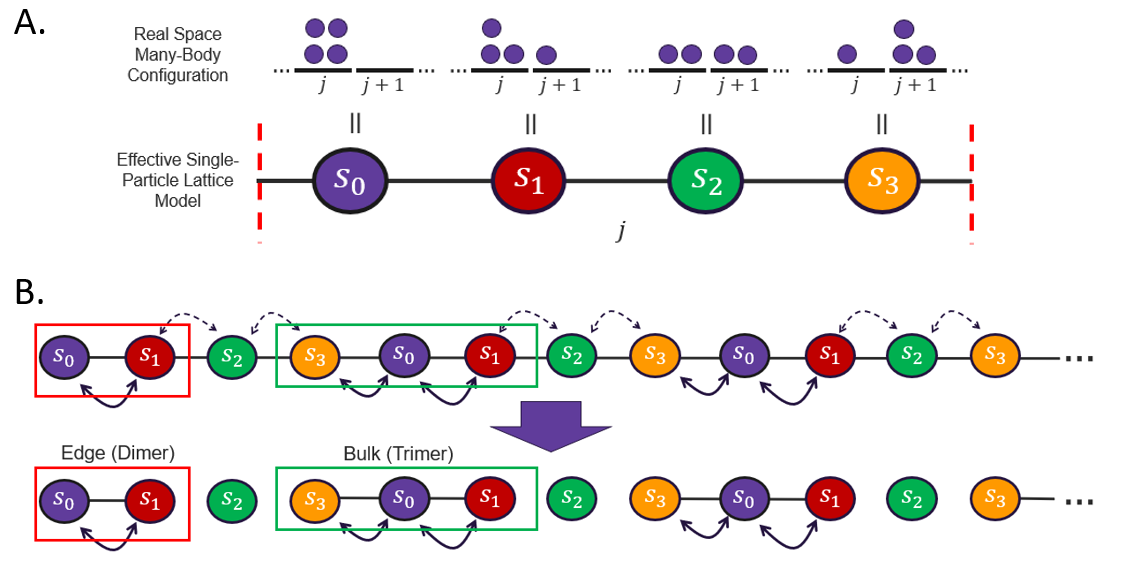}
    \caption{(A) Diagram of the real space configuration and the corresponding lattice sites of the effective Fock-space lattice model for $N=4$. (B) Diagram for $N=4$ demonstrating the relation between the full effective Fock-space lattice model and the limit where we ignore the weakest coupling that gives rise to the edge states. Ignoring the weakest coupling cuts the chain into a series of uncoupled trimers in the bulk and a dimer on each edge.}
    \label{fig:EffSSH}
\end{figure*}

An effective Fock-space lattice model can be constructed for any number of particles though it is no longer exact even in the $J=0$ limit. Still it captures the key feature of the edge states though the bulk bands no longer reflect the many-body model. For an $N$-particle system, the unit cell will contain $N$ lattice sites corresponding to partitions of the $N$ particles on two adjacent sites. We label the sites within the unit cell as $s_n$ and the unit cell with $j$. The creation operator on the $s_n$th site of the $j$th unit cell is given in real space as $a^\dagger_{s_n,j}=(a^\dagger_{j})^{N-n}(a^\dagger_{j+1})^{n}$. To make this construction more concrete, we provide the diagram of the effective model for $N=4$ particles as well as the corresponding real-space configuration for each site of the effective model in Fig.~\ref{fig:EffSSH}A. The hoppings between the $s_n$ and $s_{n+1}$ sites within the cell are given by $\sqrt{(N-n)(n+1)}(-J+\gamma(2n+1-N))$ and the intercell hopping is given by $\sqrt{N}\bigg(-J+\gamma(N-1)\bigg)$. Note there is an important point when constructing the OBC effective model, which is that the terminal site on both ends should be an $s_0$ site, i.e., all particles on the same position in the many-body picture. This is important as we will see since this gives rise to two edge states.

\begin{figure}
    \centering
    \includegraphics[width=\linewidth]{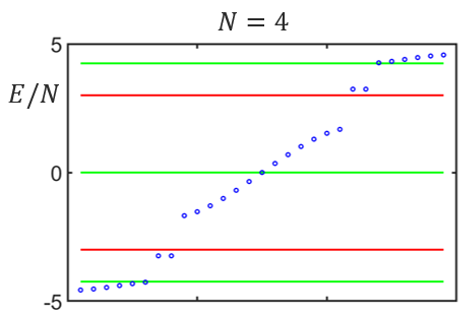}
    \caption{Spectra of the effective Fock-space lattice model for $N=4$, $L=8$. We mark the energies obtained after turning off the weakest coupling distinguishing the bulk (green lines) and edge (red lines). The horizontal axis is the eigenstate number.}
    \label{fig:DefectSpectra4}
\end{figure}

To understand the existence of the edge modes, we consider the case of $N=4$ particles. We diagram the effective lattice in the upper part of Fig.~\ref{fig:EffSSH}B. The hoppings on the diagram are given by $2(-J\mp3\gamma)$ (solid) and $\sqrt{6}(-J\mp\gamma)$ (dashed). The spectrum for this model with $J=0$, $\gamma=1$ and $N=4$, $L=8$ is presented in Fig.~\ref{fig:DefectSpectra4}, denoted by the blue circles. Here we already observe the edge modes within the band gaps. To see their origin, we set the lowest magnitude hoppings in the model, i.e $\sqrt{6}(-J\mp\gamma)$, to 0 as diagrammed in the lower part of Fig.~\ref{fig:EffSSH}B. This results in a series of uncoupled trimers in the bulk and two dimers on each edge. We diagonalize the Hamiltonian describing the trimers and dimers and mark their energies in green and red respectively on Fig.~\ref{fig:DefectSpectra4}. We see that the energy of the bulk trimers captures the bands while the edge dimers give us the energy of the midgap states. Thus, we have demonstrated that the edge states can be understood as truncation-induced states in this effective Fock-space lattice model. We emphasize that "truncation" here refers to the Fock-space lattice termination and not to the truncation of the many-body Hilbert space taken in constructing this Fock-space picture. In general, in the limit that the weakest coupling is turned off, the bulk is described by a series of uncoupled chains of length $N-1$ for $N$ even (or $N$ for an odd number of particles) while the edges are chains of length $\lceil N/2\rceil$. Turning back on the weak coupling serves to only slightly modify the spectrum. Beyond explaining the mechanism that stabilizes the edge states, this effective model provides two important pieces of information. First it allows us to determine for which values of $J$ and $\gamma$ we may be able to observe the edge mode scar. Second it allows us to efficiently determine what the most stable edge mode configuration would be by taking the highest weight $s_n$ from the edge modes obtained from the effective model. We provide a numerical demonstration of the existence of the edge modes in Appendix \ref{app:defect_edge_modes}
for $N=9$, and $N=10$ to show that this analysis works beyond 4 particles.

This reveals the presence of scar-like states consisting of $N/2$ particles bound to each edge for any system of $N$ particles, which can first be established by diagonalizing the effective Fock-space lattice model corresponding to $N/2$ particles. Once one has established the existence of the truncation-induced edge modes at a given $J$ and $\gamma$ for $N/2$ particles, they can confirm the presence of scars either through the entanglement entropy spectrum or the dynamics of the appropriate initial product state in the full many-body model with $N$ particles. As the scar is constructed as the tensor product of $N/2$ particles localized on the different edges, the frequencies observed in the fidelity oscillations result from the edges undergoing separate dynamics. For example, for the case of $N=6$ in Fig.~\ref{fig:rGamTimeEvo}, the frequency of oscillations for an individual edge is given by $\omega_\pm = |2\sqrt{3}(-J\pm2|\gamma|)|$ which is the energy cost for the particle hopping back and forth and the $\pm$ is determined by both the sign of $\gamma$ and which edge one looks at. The seemingly complicated structure of revivals in the fidelity is given by the incommensurate frequencies coming from the two edges. A Fourier analysis reveals that the component frequencies are $\omega_\pm$, $\omega_+ + \omega_-$, and $\omega_+ - \omega_-$. In addition to the results presented here, we have observed that this scarring holds for the $N=8$ case where each edge is occupied by 4 particles and for the unbalanced case where $N=7$ with 4 particles on one edge and 3 on the other. We observe that an additional particle in the bulk will collapse the localization. The energy per particle of these combined states tends to zero in the thermodynamic limit, as the interaction energy has opposite signs on each edge, preserving the QMBS interpretation of the phenomenon by pinning this state to the middle of the spectrum. In general, QMBS for other partitions of $N$ particles into $N_1$ and $N_2$ are possible and one can check the stability of the relevant edge modes in the effective Fock-space lattice models corresponding to these particle numbers to determine the fate of the scar.

\begin{figure*}
    \centering
    \includegraphics[width=\linewidth]{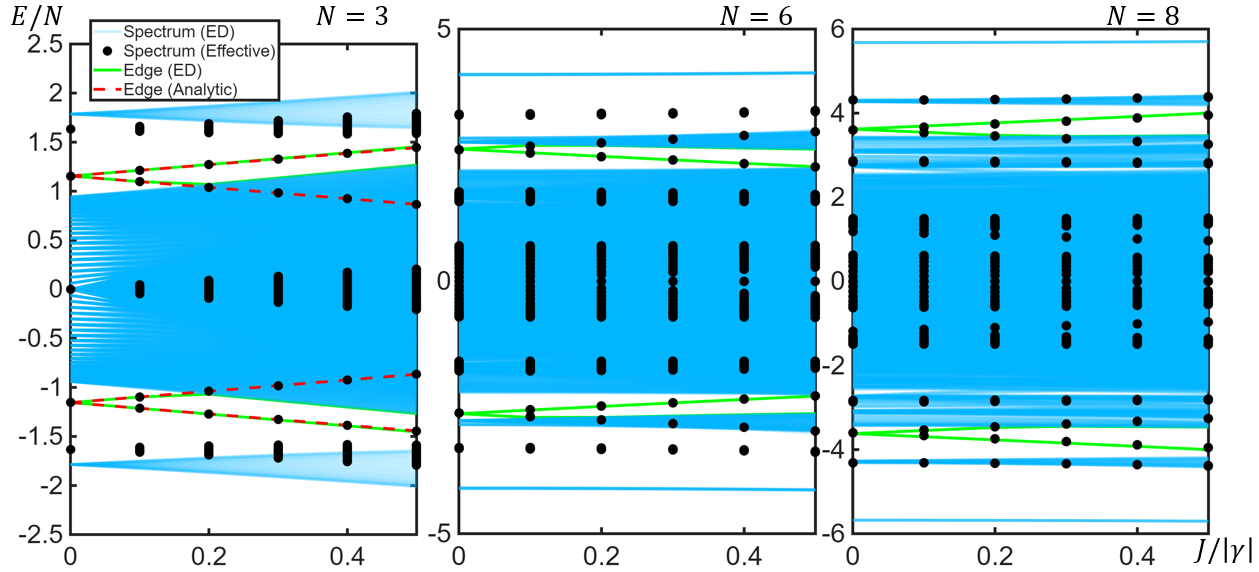}
    \caption{Spectrum as a function of $J/|\gamma|$ for particles numbers $3$, $6$, and $8$. The chain lengths are $L=12$, $12$, and $8$, respectively. The edge states are traced in green. We observe that for large $J$ they merge with the rest of the spectrum. For $N=3$, we present the analytically obtained perturbative energies for the edge modes as a red dashed line. We also include the energies obtained from the corresponding effective Fock-space lattice model as black points to demonstrate how the edge states are captured by the truncation-induced states in the effective model.}
    \label{fig:perturbative}
\end{figure*}

We further demonstrate how the scar stability weakens when the single sided edge states merge with the bound spectra. To characterize the stability of the QMBS, we calculate the spectrum for various numbers of particles fixing $\gamma = 1$, and varying $J$ from $0$ to $0.5$. At $J=0$ we observe degenerate edge states on both sides of the lattice regardless of particle number, energetically isolated from the rest of the spectrum. Turning on $J$, these states split in energy and eventually merge with extended states, which would disrupt their stability. In Fig.~\ref{fig:perturbative}, we present this calculation for $N=3, L=12$, $N=6, L=12$, and $N=8, L=8$, showing the energetically isolated edge states in green. Note that the length of the chain does not strongly affect the energies of the relevant states for this analysis. For $N=3$, the edge state is simple enough that we can analytically approximate the energy as the cost for hopping between the states $(a^\dagger_1)^3|0\rangle$ and $(a^\dagger_1)^2a^\dagger_2|0\rangle$ and the analogous process on the opposite edge. This results in branches at $\pm\sqrt{3}(-J\pm2\gamma)$, which we plot as a red dashed line in the left panel of Fig.~\ref{fig:perturbative}. We find that the critical value of $J$, where an edge state (green line) joins the continuum of extended states (blue), varies strongly with the number of particles with the values being approximately $J_c = 0.22|\gamma|$, $0.1|\gamma|$, and $0.18|\gamma|$ for $N=3$, $6$, and $8$, respectively. In general, the stability of the QMBS for a system of $N$ particles can be obtained by studying the perturbative effect of the bare hopping in the system of $N/2$ particles. We also include the energies obtained from the effective Fock-space lattice model as black points. We see that the truncation-induced states match very well with the energies of the edge states for all particle numbers, while the bulk state energies have significant inaccuracies.

\begin{figure}
    \centering
    \includegraphics[width=\linewidth]{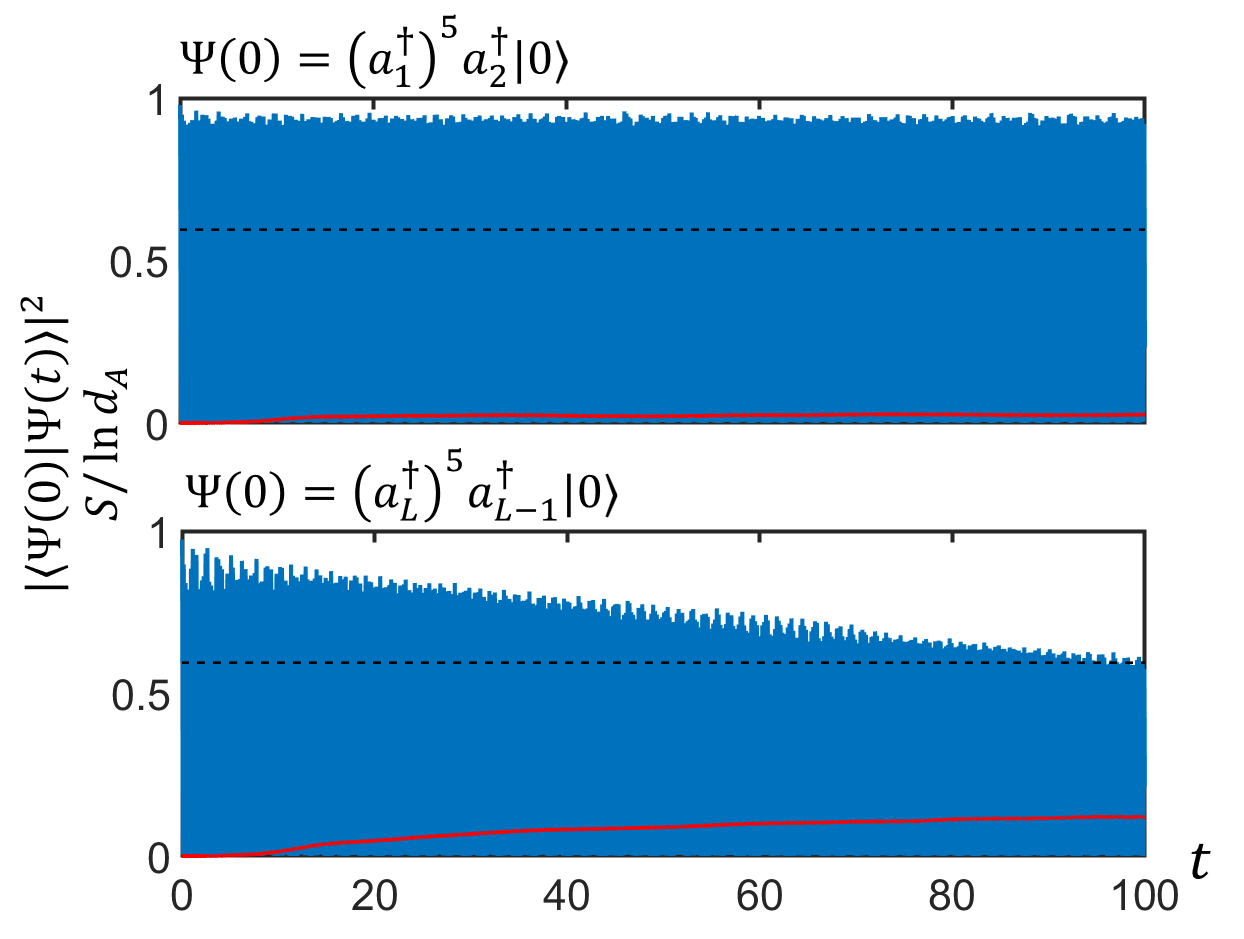}
    \caption{Time dynamics for edge mode configurations on the left (upper panel) and right (lower panel) edges for a system with $N=6$ particles and length $L=12$, $J=0.2$, $\gamma=1$. We observe that the entanglement entropy (red line) and fidelity (blue) are stable for this configuration on the left edge while the entanglement entropy grows and the fidelity decays for right edge localization.}
    \label{fig:N6stability}
\end{figure}

\begin{figure}
    \centering
    \includegraphics[width=\linewidth]{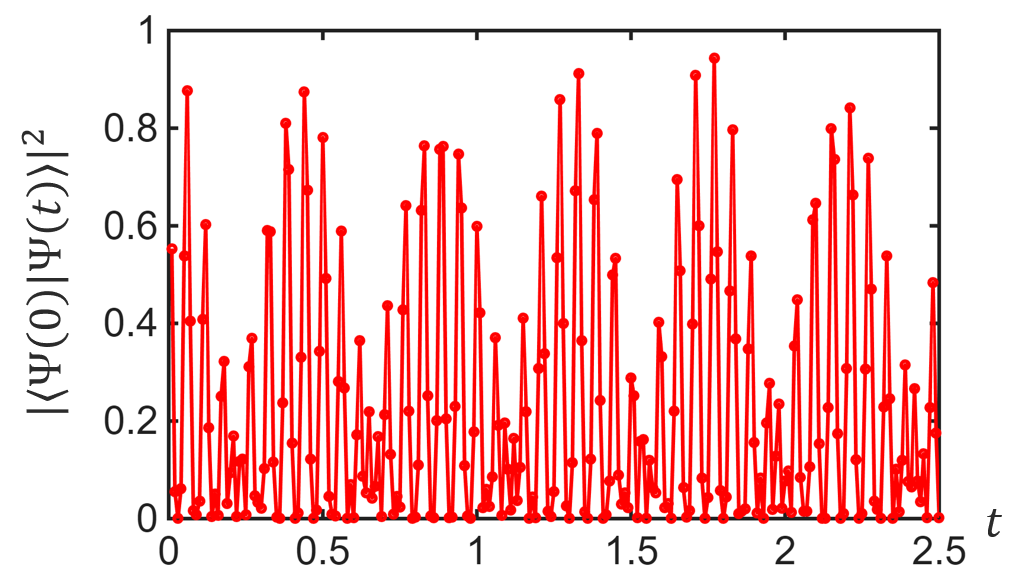}
    \caption{Time dependent fidelity obtained from TDVP for $N=10$, $L=20$, $\gamma=5$. We start from the initial state $|\Psi(0)\rangle = |410..014\rangle$ from which we observe very strong revivals in the fidelity.}
    \label{fig:TDVPreal}
\end{figure}

Finally, we demonstrate that the merging of the edge state into the continuum does indeed lower the stability of the edge mode. In Fig.~\ref{fig:N6stability}, we present the time-dependent fidelities for six particles on the left edge (upper panel) and the right edge (lower panel) in a system of $N=6$, $L=12$ for $\gamma =1$, $J= 0.2$ where we expect one edge state to lose stability due to mixing with extended states. While both edges show fast oscillations in fidelity, we see that the left edge is stable while the right shows faster growth of entanglement entropy and an overall decay in fidelity as expected from the spectrum. Still the edge state does remain stable for a significant period. We also provide TDVP calculations for a system of $N=10,L=20$ to demonstrate that the scars survive for larger systems and that we can predict the structure of the edge mode scars from our effective Fock-space lattice model. In Fig.~\ref{fig:TDVPreal}, we present the time-dependent fidelity obtained from TDVP starting from the initial state $|\Psi(0)\rangle=|410..014\rangle$ where we see the same rapid revivals as in the $N=6,L=12$ system.

\section{Discussion and Conclusion}
We have identified two forms of QMBS arising from two distinct mechanisms in the presence of density-difference-dependent hopping. First, we explored how destructive interference between the bare hopping and correlated hopping stabilizes a weak CDW scar by imposing an emergent kinetic constraint, which is absent for simple product states, but can appear for certain entangled states. This demonstrates a simple case where interference of quantum mechanical processes stabilizes otherwise unstable configurations. The mechanism we proposed also suggests the presence of weak QMBS in the presence of more general correlated hoppings. A similar CDW scar has been predicted in the context of dipolar Bose-Hubbard models~\cite{Oh24}.

Then, we explored the interplay between the formation of many-body edge states as truncation-induced edge states in an effective Fock-space lattice model and the chiral symmetry of the overall model. We found that edge-mode scars result from approximate tensor products of the edge states present for $N/2$ as evidenced by their nearly zero entanglement entropy and the frequency of oscillation. To our knowledge, these are the first example of such edge-mode QMBS. We also find that the edge-mode dynamics are remarkably sensitive to the presence of additional particles in the bulk. This QMBS demonstrates the potential role of chiral symmetry in stabilizing QMBS due to its tendency to pin states to zero energy.

QMBS have been proposed in a number of models, demonstrating that our current understanding of chaos and ergodicity in quantum mechanics is incomplete. Therefore, understanding the mechanisms by which these violations of the ETH occur is important in bridging the gap. The mechanisms proposed here are relatively simple and may provide insight into why the QMBS phenomenon appears in a variety of models, as well as how QMBS formation is disrupted in the classical limit. The use of effective single-particle models through the Krylov Hamiltonian for the CDW scar and through the cluster model for the edge scar can potentially lead to connections between QMBS and topological states. Indeed, we see this in our Krylov analysis where the symmetries of the Hamiltonian and initial state result in a Krylov Hamiltonian that is topologically nontrivial. The results presented here can also inspire investigations into QMBS of other dynamical gauge field models where similar destructive interference may occur as well as further exploring the role of symmetries in stabilizing scars.

\section*{Acknowledgments}
We would like to thank Leonardo Mazza and Yun-Tak Oh for helpful discussions. This work was funded by JSPS KAKENHI Grant Number JP24K00548, JST PRESTO Grant No. JPMJPR2353, and JST CREST Grant Number JPMJCR19T1. 
H. K. was supported by JSPS KAKENHI Grants No. JP23K25783, No. JP23K25790, and MEXT KAKENHI Grant-in-Aid for Transformative Research Areas A “Extreme Universe” (KAKENHI Grant No. JP21H05191). W. N. F. was supported by a grant from the French State, managed by the National Research Agency under the France 2030 program, reference ANR-23-PETQ-0002.

\appendix

\section{Frequency of Oscillation $J$ and $L$ Dependence}
\label{app:frequency}
\begin{figure}
    \centering
    \includegraphics[width=0.75\linewidth]{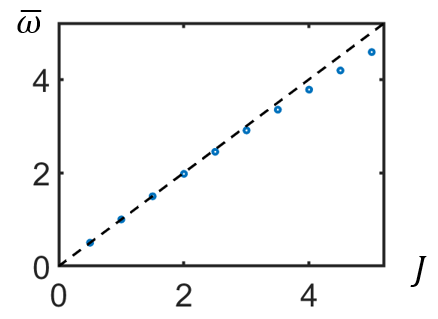}
    \caption{Average frequency ratio versus hopping $J$. These are extracted for a system of $N=6$, $L=12$, and $\gamma=10i$, by Fourier transform of the time dependent fidelity. The dashed line corresponds to $\bar\omega=J$.}
    \label{fig:JvOm}
\end{figure}

Here we provide the numerical evidence for the scaling of the frequencies mentioned in the text. To obtain the dependence of the frequencies on $J$ we fixed the coupling parameter $\gamma=10i$ and the system size to be $N=6$, $L=12$. We then extracted all oscillation frequencies through Fourier transforming the time-dependent fidelity. Finally, we take the ratio of each frequency with its corresponding partner obtained at $J=1$ and average the ratios. In Fig.~\ref{fig:JvOm}, we plot these average ratios, $\bar\omega$, versus $J$. We find that all frequencies grow proportionally with $J$ with some error as $|J|$ approaches $|\gamma|$ and the scar becomes less stable.

\begin{figure}
    \centering
    \includegraphics[width=0.75\linewidth]{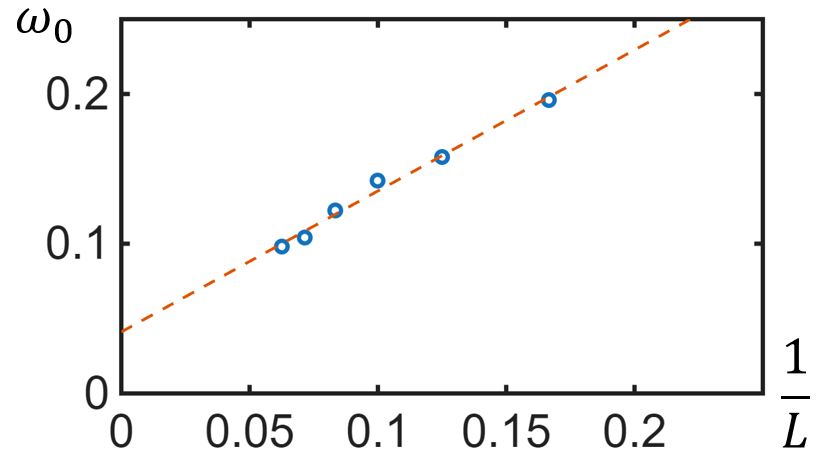}
    \caption{Fundamental frequency $\omega_0$ versus the chain length $1/L$. Markers denote frequencies obtained from the time evolution data. The dashed line is the linear fit. We see that the frequency of oscillation reduces as the system length increases, proportional to $1/L$. Here we have fixed $J=1$ and $\gamma=5i$.}
    \label{fig:LvOm}
\end{figure}

Next we consider the system size dependence of the oscillation frequency. We fix $J=1$ and $\gamma =5i$ and vary $L$ from $8$ to $16$, taking the primary lowest frequency peak from Fourier analysis. We plot the frequency vs $1/L$ in Fig.~\ref{fig:LvOm} where we see that the frequency decreases as the system size increases. This suggests that in the infinite limit the system will not oscillate between the two CDW orders, but will decohere under OBC with an even number of sites, suggesting the scar stability depends on the parity of the number of sites.

\section{Next order terms in the CDW ansatz}
\label{app:next_order}
\begin{figure}
    \centering
    \includegraphics[width=\linewidth]{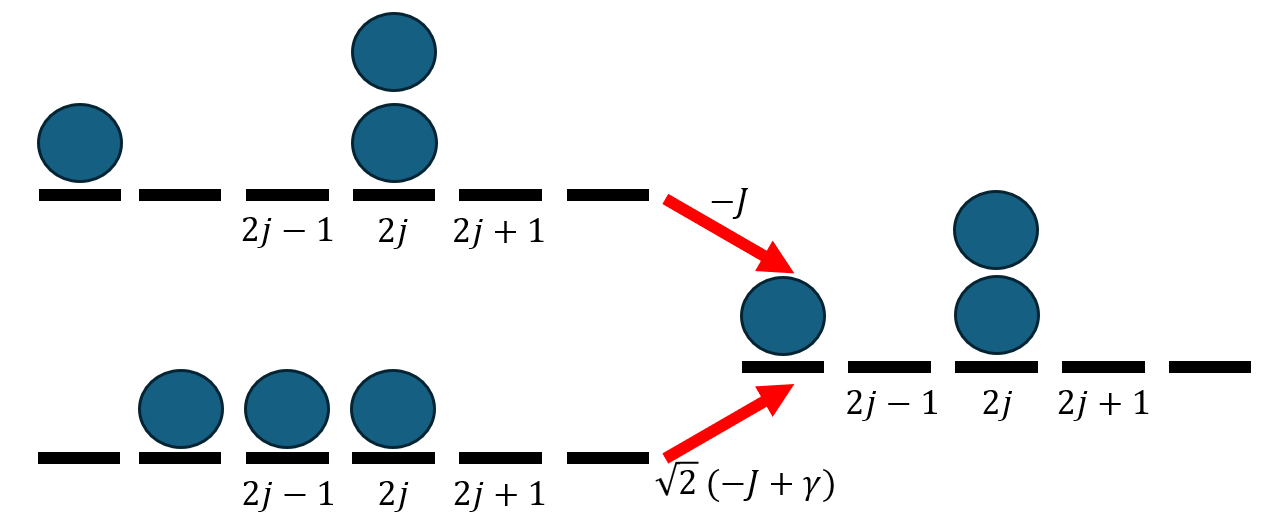}
    \caption{Diagram of the destructive interference that occurs at second order in $|J/(J+\gamma)|$.}
    \label{fig:HD2nd}
\end{figure}

Here we detail the next relevant terms to the CDW ansatz to demonstrate that they can be ignored in the limit of large $|\gamma|$. Recall that our ansatz is given by
\begin{equation}
|\Psi_{0}\rangle = c_0|{\rm CDW}\rangle - \frac{c_0J}{\sqrt{2}(J+\gamma)}\sum^{L/2}_{j=1} (a^\dagger_{2j})^2a_{2j+1}a_{2j-1}|{\rm CDW}\rangle
\end{equation}
Consider the CDW state to be the zeroth-order term and the corrections to be the first-order term. By construction, the zeroth order is frozen by destructive interference with the first order term. But the first-order term allows for two distinct classes of hopping process that are not suppressed. First, away from the bound state, the CDW order is still permitted to fluctuate. This will be suppressed by the same mechanism as in our original ansatz by interference with the unbinding of a doublon and the state mixes with configurations where there exist two bound states, naturally giving rise to an additional factor of $|J/\sqrt{2}(J+\gamma)|$ in the coefficient. The other hopping process occurs when the edge of the remaining CDW order moves toward the bound state. Due to the presence of the bound state, it is not possible to form another bound state to suppress this hopping. To achieve destructive interference, the system mixes with a configuration where three particles are adjacent. We diagram the states and the way in which they destructively interfere in Fig.~\ref{fig:HD2nd}. We see that the coefficient mixing with this state again comes with an additional factor proportional to $|J/\sqrt{2}(J+\gamma)|$, leading to a coefficient of the same order as in the case of two bound states forming far from each other. Thus we see that the ansatz we have presented is accurate in the large $\gamma$ limit.

\section{Krylov Analysis}
\label{app:kyrlov}
\begin{figure}
    \centering
    \includegraphics[width=\linewidth]{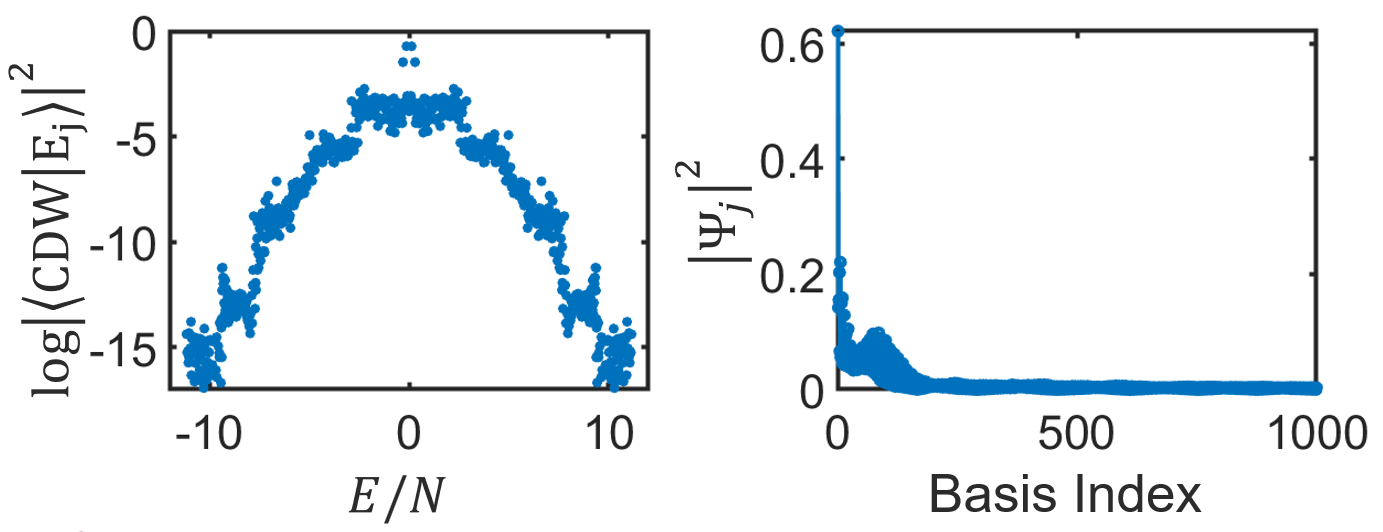}
    \caption{Left Panel: Overlaps between Krylov eigenstates and the superposition of CDW, $\frac{1}{\sqrt{2}}(|\text{CDW}\rangle+|\text{CDW'}\rangle)$. We observe large overlaps for four states near zero energy. Right panel: Density profile in Krylov space of the highest overlap state. The basis index here refers to the basis of the Krylov Hamiltonian. Note that this is an edge state in the Krylov basis suggesting the superposition of CDWs evolves slowly under time evolution. The calculation is performed for $N=6$, $L=12$, $J=1$ and $\gamma=5i$ and $1000$ iterations in the Krylov procedure.}
    \label{fig:KrylovCDW}
\end{figure}

Further evidence for the CDW state as a QMBS is provided by the topological properties of the corresponding Krylov subspace obtained as
\begin{equation}
    {\rm span}(\{H^n\frac{1}{\sqrt{2}}(|{\rm CDW}\rangle+|{\rm CDW'}\rangle), n\in\mathbb{Z} \})    
\end{equation}A similar analysis was carried out in the context of a spin model in ~\cite{Kunimi24} and we reiterate the key points below as we apply the same calculation to our model. We obtain the Krylov subspace through a Gram-Schmidt decomposition following
\begin{align}
    |u_j\rangle &= H|v_{j-1}\rangle - \sum_k^{j-1} \langle v_k|Hv_{j-1}\rangle |v_k\rangle\\
    |v_j\rangle &= \frac{|u_j\rangle}{\sqrt{\langle u_j|u_j\rangle}}
\end{align}
with $|u_0\rangle=|v_0\rangle=(|{\rm CDW}\rangle + |{\rm CDW'}\rangle)/\sqrt{2}$ where the dimension of the Krylov subspace is determined by how many iterations we perform. When rotated into this Krylov basis, the (typically truncated) Hamiltonian takes a tridiagonal form where the diagonal elements themselves are zero. We can thus treat the rotated Hamiltonian as a 1D chain with variable hopping. For $\gamma$ purely imaginary, the transformed Hamiltonian is also fully real. Additionally, if we choose an initial state that has a definite chirality (such as our CDW superposition), the Krylov Hamiltonian anticommutes with a chiral operator $\Gamma = {\rm diag}(1,-1,1,-1,...)$ and so we can treat the system as a 1D system in symmetry class BDI, permitting the existence of a topological edge mode. Note that this is the symmetry of the Krylov Hamiltonian and is not necessarily the same as the starting Hamiltonian. We calculate a Hamiltonian in the truncated Krylov space of dimension $1000$ and diagonalize to establish the existence of a localized edge mode. Such an edge mode supports the interpretation of the CDW as a scar, demonstrating that the state does not spread in time under the action of the unitary time evolution operator, $e^{-iHt}$. We present the overlaps between the superposition of CDW states and the eigenstates, $|E_j\rangle$, of the Krylov Hamiltonian, labeled by its eigenenergy $E_j$, in Fig.~\ref{fig:KrylovCDW} where we find two states, nearly degenerate in energy and overlap, with very high overlaps near zero energy. Further, we plot the profile of these states in the Krylov space in the right panel of Fig.~\ref{fig:KrylovCDW} and see that indeed these states are localized on the edge.

\section{Truncation-induced Edge modes with $N=9$ and $N=10$}
\label{app:defect_edge_modes}
\begin{figure*}
    \centering
    \includegraphics[width=0.7\linewidth]{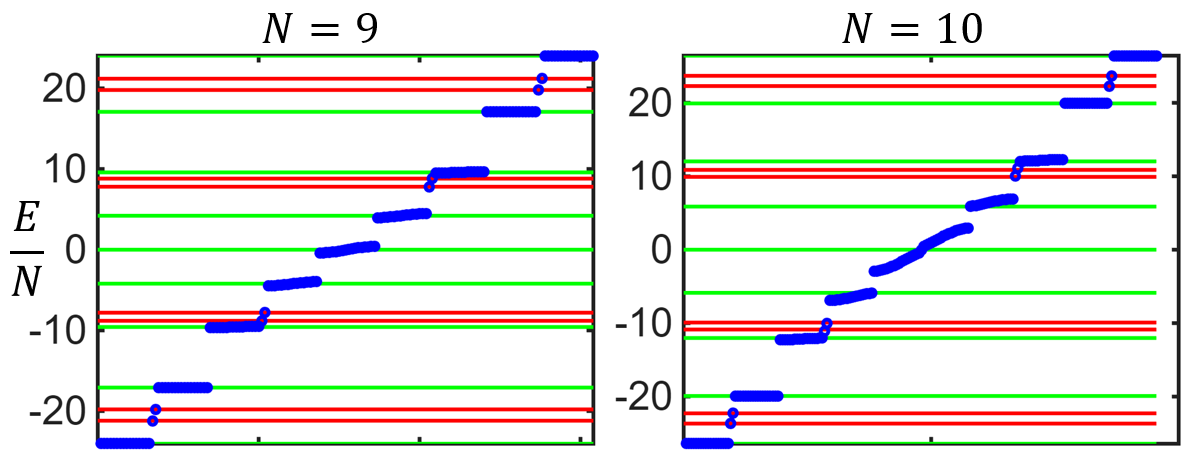}
    \caption{Spectra of the effective Fock-space lattice model for $N=9$ and $10$. We also mark the energies obtained after turning off the weakest coupling distinguishing the bulk (green lines) and edge (red lines). The horizontal axis is the eigenstate number. The parameters are chosen as $J=1,\gamma=5$.}
    \label{fig:DefectSpectra}
\end{figure*}

Here we present the numerical evidence that the truncation-induced picture of the edge mode persists for larger numbers of particles. In particular we show the numerically obtained spectra for $N=9$, $L=18$ and $N=10$, $L=20$ with $J=1$, $\gamma=5$ in the left and right panels of Fig.~\ref{fig:DefectSpectra}. We again turn off the weakest coupling and diagonalize the finite dimensional matrices corresponding to the bulk chains and the smaller edge chains.

\bibliography{biblio.bib}

@article{Atas13,
  title = {{Distribution of the Ratio of Consecutive Level Spacings in Random Matrix Ensembles}},
  author = {Atas, Y. Y. and Bogomolny, E. and Giraud, O. and Roux, G.},
  journal = {Phys. Rev. Lett.},
  volume = {110},
  issue = {8},
  pages = {084101},
  numpages = {5},
  year = {2013},
  month = {Feb},
  publisher = {American Physical Society},
  doi = {10.1103/PhysRevLett.110.084101},
  url = {https://link.aps.org/doi/10.1103/PhysRevLett.110.084101}
}

@article{Bhattacharjee22,
  title = {{Probing quantum scars and weak ergodicity breaking through quantum complexity}},
  author = {Bhattacharjee, Budhaditya and Sur, Samudra and Nandy, Pratik},
  journal = {Phys. Rev. B},
  volume = {106},
  issue = {20},
  pages = {205150},
  numpages = {15},
  year = {2022},
  month = {Nov},
  publisher = {American Physical Society},
  doi = {10.1103/PhysRevB.106.205150},
  url = {https://link.aps.org/doi/10.1103/PhysRevB.106.205150}
}

@article{Bianchi19,
  title = {{Typical entanglement entropy in the presence of a center: Page curve and its variance}},
  author = {Bianchi, Eugenio and Don\`a, Pietro},
  journal = {Phys. Rev. D},
  volume = {100},
  issue = {10},
  pages = {105010},
  numpages = {13},
  year = {2019},
  month = {Nov},
  publisher = {American Physical Society},
  doi = {10.1103/PhysRevD.100.105010},
  url = {https://link.aps.org/doi/10.1103/PhysRevD.100.105010}
}

@article{Bianchi22,
  title = {{Volume-Law Entanglement Entropy of Typical Pure Quantum States}},
  author = {Bianchi, Eugenio and Hackl, Lucas and Kieburg, Mario and Rigol, Marcos and Vidmar, Lev},
  journal = {PRX Quantum},
  volume = {3},
  issue = {3},
  pages = {030201},
  numpages = {77},
  year = {2022},
  month = {Jul},
  publisher = {American Physical Society},
  doi = {10.1103/PRXQuantum.3.030201},
  url = {https://link.aps.org/doi/10.1103/PRXQuantum.3.030201}
}

@Article{Buijsman24,
	title={{Gaussian state approximation of quantum many-body scars}},
	author={Wouter Buijsman and Yevgeny Bar Lev},
	journal={SciPost Phys.},
	volume={17},
	pages={055},
	year={2024},
	publisher={SciPost},
	doi={10.21468/SciPostPhys.17.2.055},
	url={https://scipost.org/10.21468/SciPostPhys.17.2.055},
}

@article{Chandran23,
   author = "Chandran, Anushya and Iadecola, Thomas and Khemani, Vedika and Moessner, Roderich",
   title = "Quantum Many-Body Scars: A Quasiparticle Perspective", 
   journal= "Annual Review of Condensed Matter Physics",
   year = "2023",
   volume = "14",
   number = "Volume 14, 2023",
   pages = "443-469",
   doi = "https://doi.org/10.1146/annurev-conmatphys-031620-101617",
   url = "https://www.annualreviews.org/content/journals/10.1146/annurev-conmatphys-031620-101617",
   publisher = "Annual Reviews",
   issn = "1947-5462",
   type = "Journal Article",
  }

@article{Chandran15,
  title = {{Constructing local integrals of motion in the many-body localized phase}},
  author = {Chandran, Anushya and Kim, Isaac H. and Vidal, Guifre and Abanin, Dmitry A.},
  journal = {Phys. Rev. B},
  volume = {91},
  issue = {8},
  pages = {085425},
  numpages = {7},
  year = {2015},
  month = {Feb},
  publisher = {American Physical Society},
  doi = {10.1103/PhysRevB.91.085425},
  url = {https://link.aps.org/doi/10.1103/PhysRevB.91.085425}
}

@article{DAlessio16,
author = {Luca D'Alessio, Yariv Kafri, Anatoli Polkovnikov and Marcos Rigol},
title = {{From quantum chaos and eigenstate thermalization to statistical mechanics and thermodynamics}},
journal = {Advances in Physics},
volume = {65},
number = {3},
pages = {239-362},
year = {2016},
publisher = {Taylor & Francis},
doi = {10.1080/00018732.2016.1198134},
URL = {https://doi.org/10.1080/00018732.2016.1198134},
eprint = {https://doi.org/10.1080/00018732.2016.1198134}
}

@article{DeLuca13,
doi = {10.1209/0295-5075/101/37003},
url = {https://dx.doi.org/10.1209/0295-5075/101/37003},
year = {2013},
month = {feb},
publisher = {EDP Sciences, IOP Publishing and Società Italiana di Fisica},
volume = {101},
number = {3},
pages = {37003},
author = {A. De Luca and A. Scardicchio},
title = {Ergodicity breaking in a model showing many-body localization},
journal = {Europhysics Letters}
}

@article{DeRoeck17,
  title = {Stability and instability towards delocalization in many-body localization systems},
  author = {De Roeck, Wojciech and Huveneers, Fran\ifmmode \mbox{\c{c}}\else \c{c}\fi{}ois},
  journal = {Phys. Rev. B},
  volume = {95},
  issue = {15},
  pages = {155129},
  numpages = {14},
  year = {2017},
  month = {Apr},
  publisher = {American Physical Society},
  doi = {10.1103/PhysRevB.95.155129},
  url = {https://link.aps.org/doi/10.1103/PhysRevB.95.155129}
}

@article{Deutsch91,
  title = {Quantum statistical mechanics in a closed system},
  author = {Deutsch, J. M.},
  journal = {Phys. Rev. A},
  volume = {43},
  issue = {4},
  pages = {2046--2049},
  numpages = {0},
  year = {1991},
  month = {Feb},
  publisher = {American Physical Society},
  doi = {10.1103/PhysRevA.43.2046},
  url = {https://link.aps.org/doi/10.1103/PhysRevA.43.2046}
}

@article{Evrard24,
  title = {Quantum many-body scars from unstable periodic orbits},
  author = {Evrard, Bertrand and Pizzi, Andrea and Mistakidis, Simeon I. and Dag, Ceren B.},
  journal = {Phys. Rev. B},
  volume = {110},
  issue = {14},
  pages = {144302},
  numpages = {15},
  year = {2024},
  month = {Oct},
  publisher = {American Physical Society},
  doi = {10.1103/PhysRevB.110.144302},
  url = {https://link.aps.org/doi/10.1103/PhysRevB.110.144302}
}

@article{Faugno24,
  title = {{Density Dependent Gauge Field Inducing Emergent Su-Schrieffer-Heeger Physics, Solitons, and Condensates in a Discrete Nonlinear Schr\"odinger Equation}},
  author = {Faugno, W. N. and Salerno, Mario and Ozawa, Tomoki},
  journal = {Phys. Rev. Lett.},
  volume = {132},
  issue = {2},
  pages = {023401},
  numpages = {7},
  year = {2024},
  month = {Jan},
  publisher = {American Physical Society},
  doi = {10.1103/PhysRevLett.132.023401},
  url = {https://link.aps.org/doi/10.1103/PhysRevLett.132.023401}
}

@misc{Faugno23,
      title={{Geometric Characterization of Many Body Localization}}, 
      author={W. N. Faugno and Tomoki Ozawa},
      year={2023},
      eprint={2311.12280},
      archivePrefix={arXiv},
      primaryClass={cond-mat.dis-nn}
}

@article{Frey24,
  title = {Probing Hilbert space fragmentation and the block inverse participation ratio},
  author = {Frey, Philipp and Mikhail, David and Rachel, Stephan and Hackl, Lucas},
  journal = {Phys. Rev. B},
  volume = {109},
  issue = {6},
  pages = {064302},
  numpages = {13},
  year = {2024},
  month = {Feb},
  publisher = {American Physical Society},
  doi = {10.1103/PhysRevB.109.064302},
  url = {https://link.aps.org/doi/10.1103/PhysRevB.109.064302}
}

@article{Giraud22,
  title = {{Probing Symmetries of Quantum Many-Body Systems through Gap Ratio Statistics}},
  author = {Giraud, Olivier and Mac\'e, Nicolas and Vernier, \'Eric and Alet, Fabien},
  journal = {Phys. Rev. X},
  volume = {12},
  issue = {1},
  pages = {011006},
  numpages = {25},
  year = {2022},
  month = {Jan},
  publisher = {American Physical Society},
  doi = {10.1103/PhysRevX.12.011006},
  url = {https://link.aps.org/doi/10.1103/PhysRevX.12.011006}
}

@article{Gorlach18,
  title = {Simulation of two-boson bound states using arrays of driven-dissipative coupled linear optical resonators},
  author = {Gorlach, Maxim A. and Di Liberto, Marco and Recati, Alessio and Carusotto, Iacopo and Poddubny, Alexander N. and Menotti, Chiara},
  journal = {Phys. Rev. A},
  volume = {98},
  issue = {6},
  pages = {063625},
  numpages = {11},
  year = {2018},
  month = {Dec},
  publisher = {American Physical Society},
  doi = {10.1103/PhysRevA.98.063625},
  url = {https://link.aps.org/doi/10.1103/PhysRevA.98.063625}
}

@article{Gotta23,
  title = {{Asymptotic Quantum Many-Body Scars}},
  author = {Gotta, Lorenzo and Moudgalya, Sanjay and Mazza, Leonardo},
  journal = {Phys. Rev. Lett.},
  volume = {131},
  issue = {19},
  pages = {190401},
  numpages = {7},
  year = {2023},
  month = {Nov},
  publisher = {American Physical Society},
  doi = {10.1103/PhysRevLett.131.190401},
  url = {https://link.aps.org/doi/10.1103/PhysRevLett.131.190401}
}

@article{Guo21,
  title = {Observation of energy-resolved many-body localization},
  author = {Guo, Q. and Cheng, C. and Sun, Z. H. and Song, Z. and Li, H. and Wang, Z. and Ren, W. and Dong, H. and Zheng, D. and Zhang, Y. R. and Mondaini, R. and Fan, H. and Wang, H.},
  journal = {Nature Physics},
  volume = {17},
  issue = {2},
  pages = {234-239},
  numpages = {6},
  year = {2021},
  month = {Feb},
  publisher = {Nature},
  doi = {10.1038/s41567-020-1035-1},
  url = {https://doi.org/10.1038/s41567-020-1035-1}
}

@misc{Guo23b,
      title={Observation of many-body dynamical localization}, 
      author={Yanliang Guo and Sudipta Dhar and Ang Yang and Zekai Chen and Hepeng Yao and Milena Horvath and Lei Ying and Manuele Landini and Hanns-Christoph Nägerl},
      year={2023},
      eprint={2312.13880},
      archivePrefix={arXiv},
      primaryClass={quant-ph},
      url={https://arxiv.org/abs/2312.13880}, 
}

@article{Huang21b,
  title = {{Stability of scar states in the two-dimensional PXP model against random disorder}},
  author = {Huang, Ke and Wang, Yu and Li, Xiao},
  journal = {Phys. Rev. B},
  volume = {104},
  issue = {21},
  pages = {214305},
  numpages = {10},
  year = {2021},
  month = {Dec},
  publisher = {American Physical Society},
  doi = {10.1103/PhysRevB.104.214305},
  url = {https://link.aps.org/doi/10.1103/PhysRevB.104.214305}
}

@article{Hudomal20,
  title = {Quantum scars of bosons with correlated hopping},
  author = {A. Hudomal and I. Vasić and N. Regnault and Z. Papić},
  journal = {Communications Physics},
  volume = {3},
  issue = {1},
  pages = {2399-3650},
  year = {2020},
  month = {Jun},
  publisher = {Nature},
  doi = {10.1038/s42005-020-0364-9},
  url = {https://doi.org/10.1038/s42005-020-0364-9}
}

@article{Hudomal22,
  title = {{Driving quantum many-body scars in the PXP model}},
  author = {Hudomal, Ana and Desaules, Jean-Yves and Mukherjee, Bhaskar and Su, Guo-Xian and Halimeh, Jad C. and Papi\ifmmode \acute{c}\else \'{c}\fi{}, Zlatko},
  journal = {Phys. Rev. B},
  volume = {106},
  issue = {10},
  pages = {104302},
  numpages = {19},
  year = {2022},
  month = {Sep},
  publisher = {American Physical Society},
  doi = {10.1103/PhysRevB.106.104302},
  url = {https://link.aps.org/doi/10.1103/PhysRevB.106.104302}
}

@article{Hummel23,
  title = {{Genuine Many-Body Quantum Scars along Unstable Modes in Bose-Hubbard Systems}},
  author = {Hummel, Quirin and Richter, Klaus and Schlagheck, Peter},
  journal = {Phys. Rev. Lett.},
  volume = {130},
  issue = {25},
  pages = {250402},
  numpages = {6},
  year = {2023},
  month = {Jun},
  publisher = {American Physical Society},
  doi = {10.1103/PhysRevLett.130.250402},
  url = {https://link.aps.org/doi/10.1103/PhysRevLett.130.250402}
}

@article{Huse14,
  title = {Phenomenology of fully many-body-localized systems},
  author = {Huse, David A. and Nandkishore, Rahul and Oganesyan, Vadim},
  journal = {Phys. Rev. B},
  volume = {90},
  issue = {17},
  pages = {174202},
  numpages = {5},
  year = {2014},
  month = {Nov},
  publisher = {American Physical Society},
  doi = {10.1103/PhysRevB.90.174202},
  url = {https://link.aps.org/doi/10.1103/PhysRevB.90.174202}
}

@article{Jensen85,
  title = {{Statistical Behavior in Deterministic Quantum Systems with Few Degrees of Freedom}},
  author = {Jensen, R. V. and Shankar, R.},
  journal = {Phys. Rev. Lett.},
  volume = {54},
  issue = {17},
  pages = {1879--1882},
  numpages = {0},
  year = {1985},
  month = {Apr},
  publisher = {American Physical Society},
  doi = {10.1103/PhysRevLett.54.1879},
  url = {https://link.aps.org/doi/10.1103/PhysRevLett.54.1879}
}

@article{Keneko24,
  title = {{Quantum many-body scars in the Bose-Hubbard model with a three-body constraint}},
  author = {Kaneko, Ryui and Kunimi, Masaya and Danshita, Ippei},
  journal = {Phys. Rev. A},
  volume = {109},
  issue = {1},
  pages = {L011301},
  numpages = {6},
  year = {2024},
  month = {Jan},
  publisher = {American Physical Society},
  doi = {10.1103/PhysRevA.109.L011301},
  url = {https://link.aps.org/doi/10.1103/PhysRevA.109.L011301}
}

@article{Kohlert19,
  title = {{Observation of Many-Body Localization in a One-Dimensional System with a Single-Particle Mobility Edge}},
  author = {Kohlert, Thomas and Scherg, Sebastian and Li, Xiao and L\"uschen, Henrik P. and Das Sarma, Sankar and Bloch, Immanuel and Aidelsburger, Monika},
  journal = {Phys. Rev. Lett.},
  volume = {122},
  issue = {17},
  pages = {170403},
  numpages = {7},
  year = {2019},
  month = {May},
  publisher = {American Physical Society},
  doi = {10.1103/PhysRevLett.122.170403},
  url = {https://link.aps.org/doi/10.1103/PhysRevLett.122.170403}
}

@article{Kunimi24,
  title = {{Proposal for simulating quantum spin models with the Dzyaloshinskii-Moriya interaction using Rydberg atoms and the construction of asymptotic quantum many-body scar states}},
  author = {Kunimi, Masaya and Tomita, Takafumi and Katsura, Hosho and Kato, Yusuke},
  journal = {Phys. Rev. A},
  volume = {110},
  issue = {4},
  pages = {043312},
  numpages = {18},
  year = {2024},
  month = {Oct},
  publisher = {American Physical Society},
  doi = {10.1103/PhysRevA.110.043312},
  url = {https://link.aps.org/doi/10.1103/PhysRevA.110.043312}
}

@article{Lake22,
  title = {{Dipolar Bose-Hubbard model}},
  author = {Lake, Ethan and Hermele, Michael and Senthil, T.},
  journal = {Phys. Rev. B},
  volume = {106},
  issue = {6},
  pages = {064511},
  numpages = {14},
  year = {2022},
  month = {Aug},
  publisher = {American Physical Society},
  doi = {10.1103/PhysRevB.106.064511},
  url = {https://link.aps.org/doi/10.1103/PhysRevB.106.064511}
}

@article{Lake23,
  title = {{Dipole condensates in tilted Bose-Hubbard chains}},
  author = {Lake, Ethan and Lee, Hyun-Yong and Han, Jung Hoon and Senthil, T.},
  journal = {Phys. Rev. B},
  volume = {107},
  issue = {19},
  pages = {195132},
  numpages = {20},
  year = {2023},
  month = {May},
  publisher = {American Physical Society},
  doi = {10.1103/PhysRevB.107.195132},
  url = {https://link.aps.org/doi/10.1103/PhysRevB.107.195132}
}

@article{Lan18,
  title = {{Quantum Slow Relaxation and Metastability due to Dynamical Constraints}},
  author = {Lan, Zhihao and van Horssen, Merlijn and Powell, Stephen and Garrahan, Juan P.},
  journal = {Phys. Rev. Lett.},
  volume = {121},
  issue = {4},
  pages = {040603},
  numpages = {6},
  year = {2018},
  month = {Jul},
  publisher = {American Physical Society},
  doi = {10.1103/PhysRevLett.121.040603},
  url = {https://link.aps.org/doi/10.1103/PhysRevLett.121.040603}
}

@article{Levi16,
  title = {{Robustness of Many-Body Localization in the Presence of Dissipation}},
  author = {Levi, Emanuele and Heyl, Markus and Lesanovsky, Igor and Garrahan, Juan P.},
  journal = {Phys. Rev. Lett.},
  volume = {116},
  issue = {23},
  pages = {237203},
  numpages = {5},
  year = {2016},
  month = {Jun},
  publisher = {American Physical Society},
  doi = {10.1103/PhysRevLett.116.237203},
  url = {https://link.aps.org/doi/10.1103/PhysRevLett.116.237203}
}

@article{Ljubotina23,
  title = {{Superdiffusive Energy Transport in Kinetically Constrained Models}},
  author = {Ljubotina, Marko and Desaules, Jean-Yves and Serbyn, Maksym and Papi\ifmmode \acute{c}\else \'{c}\fi{}, Zlatko},
  journal = {Phys. Rev. X},
  volume = {13},
  issue = {1},
  pages = {011033},
  numpages = {14},
  year = {2023},
  month = {Mar},
  publisher = {American Physical Society},
  doi = {10.1103/PhysRevX.13.011033},
  url = {https://link.aps.org/doi/10.1103/PhysRevX.13.011033}
}

@article{McClarty20,
  title = {Disorder-free localization and many-body quantum scars from magnetic frustration},
  author = {McClarty, Paul A. and Haque, Masudul and Sen, Arnab and Richter, Johannes},
  journal = {Phys. Rev. B},
  volume = {102},
  issue = {22},
  pages = {224303},
  numpages = {15},
  year = {2020},
  month = {Dec},
  publisher = {American Physical Society},
  doi = {10.1103/PhysRevB.102.224303},
  url = {https://link.aps.org/doi/10.1103/PhysRevB.102.224303}
}

@article{Morong21,
  title = {{Observation of Stark many-body localization without disorder}},
  author = {Morong, W. and Liu, F. and Becker, P. and Collins, K. S. and Feng, L. and Kyprianidis, A. and Pagano, G. and You, T. and Gorshkov, A. V. and Monroe, C.},
  journal = {Nature},
  volume = {599},
  issue = {7885},
  pages = {393-398},
  numpages = {6},
  year = {2021},
  month = {Jan},
  publisher = {Nature},
  doi = {10.1038/s41586-021-03988-0},
  url = {https://doi.org/10.1038/s41586-021-03988-0}
}

@article{Moudgalya22,
doi = {10.1088/1361-6633/ac73a0},
url = {https://dx.doi.org/10.1088/1361-6633/ac73a0},
year = {2022},
month = {jul},
publisher = {IOP Publishing},
volume = {85},
number = {8},
pages = {086501},
author = {Sanjay Moudgalya and B Andrei Bernevig and Nicolas Regnault},
title = {{Quantum many-body scars and Hilbert space fragmentation: a review of exact results}},
journal = {Reports on Progress in Physics}
}

@article{Moudgalya20,
  title = {{Quantum many-body scars in a Landau level on a thin torus}},
  author = {Moudgalya, Sanjay and Bernevig, B. Andrei and Regnault, Nicolas},
  journal = {Phys. Rev. B},
  volume = {102},
  issue = {19},
  pages = {195150},
  numpages = {27},
  year = {2020},
  month = {Nov},
  publisher = {American Physical Society},
  doi = {10.1103/PhysRevB.102.195150},
  url = {https://link.aps.org/doi/10.1103/PhysRevB.102.195150}
}

@article{Nandy24,
doi = {10.1088/1361-648X/ad1a7b},
url = {https://dx.doi.org/10.1088/1361-648X/ad1a7b},
year = {2024},
month = {jan},
publisher = {IOP Publishing},
volume = {36},
number = {15},
pages = {155601},
author = {Nandy, Sourav and Mukherjee, Bhaskar and Bhattacharyya, Arpan and Banerjee, Aritra},
title = {Quantum state complexity meets many-body scars},
journal = {Journal of Physics: Condensed Matter}
}

@article{Oh24,
  title = {{Fractonic quantum quench in dipole-constrained bosons}},
  author = {Oh, Yun-Tak and Han, Jung Hoon and Lee, Hyun-Yong},
  journal = {Phys. Rev. Res.},
  volume = {6},
  issue = {2},
  pages = {023269},
  numpages = {11},
  year = {2024},
  month = {Jun},
  publisher = {American Physical Society},
  doi = {10.1103/PhysRevResearch.6.023269},
  url = {https://link.aps.org/doi/10.1103/PhysRevResearch.6.023269}
}

@article{Omiya23,
  title = {{Quantum many-body scars in bipartite Rydberg arrays originating from hidden projector embedding}},
  author = {Omiya, Keita and M\"uller, Markus},
  journal = {Phys. Rev. A},
  volume = {107},
  issue = {2},
  pages = {023318},
  numpages = {20},
  year = {2023},
  month = {Feb},
  publisher = {American Physical Society},
  doi = {10.1103/PhysRevA.107.023318},
  url = {https://link.aps.org/doi/10.1103/PhysRevA.107.023318}
}

@article{Page93,
  title = {Average entropy of a subsystem},
  author = {Page, Don N.},
  journal = {Phys. Rev. Lett.},
  volume = {71},
  issue = {9},
  pages = {1291--1294},
  numpages = {0},
  year = {1993},
  month = {Aug},
  publisher = {American Physical Society},
  doi = {10.1103/PhysRevLett.71.1291},
  url = {https://link.aps.org/doi/10.1103/PhysRevLett.71.1291}
}

@article{Pal10,
  title = {Many-body localization phase transition},
  author = {Pal, Arijeet and Huse, David A.},
  journal = {Phys. Rev. B},
  volume = {82},
  issue = {17},
  pages = {174411},
  numpages = {7},
  year = {2010},
  month = {Nov},
  publisher = {American Physical Society},
  doi = {10.1103/PhysRevB.82.174411},
  url = {https://link.aps.org/doi/10.1103/PhysRevB.82.174411}
}

@article{Pancotti20,
  title = {{Quantum East Model: Localization, Nonthermal Eigenstates, and Slow Dynamics}},
  author = {Pancotti, Nicola and Giudice, Giacomo and Cirac, J. Ignacio and Garrahan, Juan P. and Ba\~nuls, Mari Carmen},
  journal = {Phys. Rev. X},
  volume = {10},
  issue = {2},
  pages = {021051},
  numpages = {21},
  year = {2020},
  month = {Jun},
  publisher = {American Physical Society},
  doi = {10.1103/PhysRevX.10.021051},
  url = {https://link.aps.org/doi/10.1103/PhysRevX.10.021051}
}

@article{Pekker17,
  title = {{Fixed Points of Wegner-Wilson Flows and Many-Body Localization}},
  author = {Pekker, David and Clark, Bryan K. and Oganesyan, Vadim and Refael, Gil},
  journal = {Phys. Rev. Lett.},
  volume = {119},
  issue = {7},
  pages = {075701},
  numpages = {5},
  year = {2017},
  month = {Aug},
  publisher = {American Physical Society},
  doi = {10.1103/PhysRevLett.119.075701},
  url = {https://link.aps.org/doi/10.1103/PhysRevLett.119.075701}
}

@article{Pinto09,
  title = {Edge-localized states in quantum one-dimensional lattices},
  author = {Pinto, Ricardo A. and Haque, Masudul and Flach, Sergej},
  journal = {Phys. Rev. A},
  volume = {79},
  issue = {5},
  pages = {052118},
  numpages = {8},
  year = {2009},
  month = {May},
  publisher = {American Physical Society},
  doi = {10.1103/PhysRevA.79.052118},
  url = {https://link.aps.org/doi/10.1103/PhysRevA.79.052118}
}

@article{Rademaker16,
  title = {{Explicit Local Integrals of Motion for the Many-Body Localized State}},
  author = {Rademaker, Louk and Ortu\~no, Miguel},
  journal = {Phys. Rev. Lett.},
  volume = {116},
  issue = {1},
  pages = {010404},
  numpages = {5},
  year = {2016},
  month = {Jan},
  publisher = {American Physical Society},
  doi = {10.1103/PhysRevLett.116.010404},
  url = {https://link.aps.org/doi/10.1103/PhysRevLett.116.010404}
}

@article{Rigol08,
  title = {Thermalization and its mechanism for generic isolated quantum systems},
  author = {Rigol, Marcos and Dunjko, Vanja and Olshanii, Maxim},
  journal = {Nature},
  volume = {452},
  issue = {7189},
  pages = {854858},
  numpages = {4},
  year = {2008},
  month = {Apr},
  publisher = {Nature},
  doi = {10.1038/nature06838},
  url = {https://doi.org/10.1038/nature06838}
}

@article{Sala20,
  title = {{Ergodicity Breaking Arising from Hilbert Space Fragmentation in Dipole-Conserving Hamiltonians}},
  author = {Sala, Pablo and Rakovszky, Tibor and Verresen, Ruben and Knap, Michael and Pollmann, Frank},
  journal = {Phys. Rev. X},
  volume = {10},
  issue = {1},
  pages = {011047},
  numpages = {19},
  year = {2020},
  month = {Feb},
  publisher = {American Physical Society},
  doi = {10.1103/PhysRevX.10.011047},
  url = {https://link.aps.org/doi/10.1103/PhysRevX.10.011047}
}

@article{Sanada23,
  title = {Quantum many-body scars in spin models with multibody interactions},
  author = {Sanada, Kazuyuki and Miao, Yuan and Katsura, Hosho},
  journal = {Phys. Rev. B},
  volume = {108},
  issue = {15},
  pages = {155102},
  numpages = {25},
  year = {2023},
  month = {Oct},
  publisher = {American Physical Society},
  doi = {10.1103/PhysRevB.108.155102},
  url = {https://link.aps.org/doi/10.1103/PhysRevB.108.155102}
}

@article{Serbyn21,
  title = {Quantum many-body scars and weak breaking of ergodicity},
  author = {Serbyn, Maksym and Abanin, Dmitry A. and Papi\ifmmode \acute{c}\else \'{c}\fi{}, Z. },
  journal = {Nature Physics},
  volume = {17},
  issue = {6},
  pages = {675685},
  numpages = {10},
  year = {2021},
  month = {Jun},
  publisher = {Nature},
  doi = {10.1038/s41567-021-01230-2},
  url = {https://doi.org/10.1038/s41567-021-01230-2}
}

@article{Serbyn13b,
  title = {{Local Conservation Laws and the Structure of the Many-Body Localized States}},
  author = {Serbyn, Maksym and Papi\ifmmode \acute{c}\else \'{c}\fi{}, Z. and Abanin, Dmitry A.},
  journal = {Phys. Rev. Lett.},
  volume = {111},
  issue = {12},
  pages = {127201},
  numpages = {5},
  year = {2013},
  month = {Sep},
  publisher = {American Physical Society},
  doi = {10.1103/PhysRevLett.111.127201},
  url = {https://link.aps.org/doi/10.1103/PhysRevLett.111.127201}
}

@article{Srednicki94,
  title = {Chaos and quantum thermalization},
  author = {Srednicki, Mark},
  journal = {Phys. Rev. E},
  volume = {50},
  issue = {2},
  pages = {888--901},
  numpages = {0},
  year = {1994},
  month = {Aug},
  publisher = {American Physical Society},
  doi = {10.1103/PhysRevE.50.888},
  url = {https://link.aps.org/doi/10.1103/PhysRevE.50.888}
}

@article{Su23,
  title = {{Observation of many-body scarring in a Bose-Hubbard quantum simulator}},
  author = {Su, Guo-Xian and Sun, Hui and Hudomal, Ana and Desaules, Jean-Yves and Zhou, Zhao-Yu and Yang, Bing and Halimeh, Jad C. and Yuan, Zhen-Sheng and Papi\ifmmode \acute{c}\else \'{c}\fi{}, Zlatko and Pan, Jian-Wei},
  journal = {Phys. Rev. Res.},
  volume = {5},
  issue = {2},
  pages = {023010},
  numpages = {13},
  year = {2023},
  month = {Apr},
  publisher = {American Physical Society},
  doi = {10.1103/PhysRevResearch.5.023010},
  url = {https://link.aps.org/doi/10.1103/PhysRevResearch.5.023010}
}

@article{Tamura22,
  title = {Quantum many-body scars of spinless fermions with density-assisted hopping in higher dimensions},
  author = {Tamura, Kensuke and Katsura, Hosho},
  journal = {Phys. Rev. B},
  volume = {106},
  issue = {14},
  pages = {144306},
  numpages = {11},
  year = {2022},
  month = {Oct},
  publisher = {American Physical Society},
  doi = {10.1103/PhysRevB.106.144306},
  url = {https://link.aps.org/doi/10.1103/PhysRevB.106.144306}
}

@article{Theel25,
  title = {Chirally Protected State Manipulation by Tuning One-Dimensional Statistics},
  author = {Theel, F. and Bonkhoff, M. and Schmelcher, P. and Posske, T. and Harshman, N. L.},
  journal = {Phys. Rev. Lett.},
  volume = {135},
  issue = {6},
  pages = {063401},
  numpages = {6},
  year = {2025},
  month = {Aug},
  publisher = {American Physical Society},
  doi = {10.1103/kzf6-yx24},
  url = {https://link.aps.org/doi/10.1103/kzf6-yx24}
}

@article{Thiery18,
  title = {{Many-Body Delocalization as a Quantum Avalanche}},
  author = {Thiery, Thimoth\'ee and Huveneers, Fran\ifmmode \mbox{\c{c}}\else \c{c}\fi{}ois and M\"uller, Markus and De Roeck, Wojciech},
  journal = {Phys. Rev. Lett.},
  volume = {121},
  issue = {14},
  pages = {140601},
  numpages = {6},
  year = {2018},
  month = {Oct},
  publisher = {American Physical Society},
  doi = {10.1103/PhysRevLett.121.140601},
  url = {https://link.aps.org/doi/10.1103/PhysRevLett.121.140601}
}

@article{Thomson18,
  title = {Time evolution of many-body localized systems with the flow equation approach},
  author = {Thomson, S. J. and Schir\'o, M.},
  journal = {Phys. Rev. B},
  volume = {97},
  issue = {6},
  pages = {060201},
  numpages = {5},
  year = {2018},
  month = {Feb},
  publisher = {American Physical Society},
  doi = {10.1103/PhysRevB.97.060201},
  url = {https://link.aps.org/doi/10.1103/PhysRevB.97.060201}
}

@article{Turner21,
  title = {{Correspondence Principle for Many-Body Scars in Ultracold Rydberg Atoms}},
  author = {Turner, C. J. and Desaules, J.-Y. and Bull, K. and Papi\ifmmode \acute{c}\else \'{c}\fi{}, Z.},
  journal = {Phys. Rev. X},
  volume = {11},
  issue = {2},
  pages = {021021},
  numpages = {22},
  year = {2021},
  month = {Apr},
  publisher = {American Physical Society},
  doi = {10.1103/PhysRevX.11.021021},
  url = {https://link.aps.org/doi/10.1103/PhysRevX.11.021021}
}

@article{Turner18b,
  title = {{Quantum scarred eigenstates in a Rydberg atom chain: Entanglement, breakdown of thermalization, and stability to perturbations}},
  author = {Turner, C. J. and Michailidis, A. A. and Abanin, D. A. and Serbyn, M. and Papi\ifmmode \acute{c}\else \'{c}\fi{}, Z.},
  journal = {Phys. Rev. B},
  volume = {98},
  issue = {15},
  pages = {155134},
  numpages = {23},
  year = {2018},
  month = {Oct},
  publisher = {American Physical Society},
  doi = {10.1103/PhysRevB.98.155134},
  url = {https://link.aps.org/doi/10.1103/PhysRevB.98.155134}
}

@article{Turner18a,
  title = {Weak ergodicity breaking from quantum many-body scars},
  author = {Turner, C. J. and Michailidis, A. A. and Abanin, D. A. and Serbyn, M. and Papi\ifmmode \acute{c}\else \'{c}\fi{}, Z.},
  journal = {Nature Physics},
  volume = {14},
  issue = {7},
  pages = {745749},
  numpages = {5},
  year = {2018},
  month = {May},
  publisher = {Nature},
  doi = {10.1038/s41567-018-0137-5},
  url = {https://doi.org/10.1038/s41567-018-0137-5}
}

@article{Turner18,
  title = {{Quantum scarred eigenstates in a Rydberg atom chain: Entanglement, breakdown of thermalization, and stability to perturbations}},
  author = {Turner, C. J. and Michailidis, A. A. and Abanin, D. A. and Serbyn, M. and Papi\ifmmode \acute{c}\else \'{c}\fi{}, Z.},
  journal = {Phys. Rev. B},
  volume = {98},
  issue = {15},
  pages = {155134},
  numpages = {23},
  year = {2018},
  month = {Oct},
  publisher = {American Physical Society},
  doi = {10.1103/PhysRevB.98.155134},
  url = {https://link.aps.org/doi/10.1103/PhysRevB.98.155134}
}

@article{Vidmar17,
  title = {{Entanglement Entropy of Eigenstates of Quantum Chaotic Hamiltonians}},
  author = {Vidmar, Lev and Rigol, Marcos},
  journal = {Phys. Rev. Lett.},
  volume = {119},
  issue = {22},
  pages = {220603},
  numpages = {6},
  year = {2017},
  month = {Nov},
  publisher = {American Physical Society},
  doi = {10.1103/PhysRevLett.119.220603},
  url = {https://link.aps.org/doi/10.1103/PhysRevLett.119.220603}
}

@article{Yang20,
  title = {{Hilbert-Space Fragmentation from Strict Confinement}},
  author = {Yang, Zhi-Cheng and Liu, Fangli and Gorshkov, Alexey V. and Iadecola, Thomas},
  journal = {Phys. Rev. Lett.},
  volume = {124},
  issue = {20},
  pages = {207602},
  numpages = {6},
  year = {2020},
  month = {May},
  publisher = {American Physical Society},
  doi = {10.1103/PhysRevLett.124.207602},
  url = {https://link.aps.org/doi/10.1103/PhysRevLett.124.207602}
}

@article{gotta2022exact,
  title={Exact many-body scars based on pairs or multimers in a chain of spinless fermions},
  author={Gotta, Lorenzo and Mazza, Leonardo and Simon, Pascal and Roux, Guillaume},
  journal={Physical Review B},
  volume={106},
  number={23},
  pages={235147},
  year={2022},
  publisher={APS},
  doi={10.1103/PhysRevB.106.235147},
  url={https://doi.org/10.1103/PhysRevB.106.235147}
}

@Article{10.21468/SciPostPhysCodeb.4,
	title={{The ITensor Software Library for Tensor Network Calculations}},
	author={Matthew Fishman and Steven R. White and E. Miles Stoudenmire},
	journal={SciPost Phys. Codebases},
	pages={4},
	year={2022},
	publisher={SciPost},
	doi={10.21468/SciPostPhysCodeb.4},
	url={https://scipost.org/10.21468/SciPostPhysCodeb.4},
}

@Article{10.21468/SciPostPhysCodeb.4-r0.3,
	title={{Codebase release 0.3 for ITensor}},
	author={Matthew Fishman and Steven R. White and E. Miles Stoudenmire},
	journal={SciPost Phys. Codebases},
	pages={4-r0.3},
	year={2022},
	publisher={SciPost},
	doi={10.21468/SciPostPhysCodeb.4-r0.3},
	url={https://scipost.org/10.21468/SciPostPhysCodeb.4-r0.3},
}
\end{document}